\documentclass[journal=jctcce,manuscript=article,email=false]{achemso}
\setkeys{acs}{articletitle=true,etalmode=truncate,maxauthors=0}

\usepackage[colorlinks=true,linkcolor=blue,citecolor=blue,urlcolor=blue]{hyperref}
\usepackage{setspace}
\usepackage{graphicx}
\usepackage{amsmath}
\usepackage{color}
\usepackage{amssymb}
\usepackage{verbatim}
\usepackage{latexsym}
\usepackage{enumerate}
\usepackage{bm}
\usepackage{caption}
\usepackage{subcaption}
\usepackage{mathtools}
\usepackage[utf8]{inputenc}
\usepackage[T1]{fontenc}
\usepackage{mciteplus}
\usepackage{xr}

\title{Accurate and efficient computation of optical absorption
  spectra of molecular crystals: the case of the polymorphs of ROY}

\author{Joseph C.\ A.\ Prentice}
\affiliation{Department of Materials, University of Oxford, Parks
  Road, Oxford OX1 3PH, United Kingdom}
\alsoaffiliation{Department of Materials, Department of Physics, and the
  Thomas Young Centre for Theory and Simulation of Materials, Imperial
  College London, London SW7 2AZ, United Kingdom}
\author{Arash A.\ Mostofi}
\affiliation{Department of Materials, Department of Physics, and the
  Thomas Young Centre for Theory and Simulation of Materials, Imperial
  College London, London SW7 2AZ, United Kingdom}

\begin{document}

\begin{abstract}

  When calculating the optical absorption spectra of molecular
  crystals from first principles, the influence of the crystalline
  environment on the excitations is of significant importance. For
  such systems, however, methods to describe the excitations
  accurately can be computationally prohibitive due to the relatively
  large system sizes involved. In this work, we demonstrate a method
  that allows optical absorption spectra to be computed both
  efficiently and at high accuracy. Our approach is based on the
  spectral warping method successfully applied to molecules in
  solvent. It involves calculating the absorption spectrum of a
  supercell of the full molecular crystal using semi-local TDDFT,
  before warping the spectrum using a transformation derived from
  smaller-scale semi-local and hybrid TDDFT calculations on isolated
  dimers. We demonstrate the power of this method on three polymorphs
  of the well known colour polymorphic compound ROY, and find that it
  outperforms both small-scale hybrid TDDFT dimer calculations and
  large-scale semi-local TDDFT supercell calculations, when compared
  to experiment.

\end{abstract}

\section{Introduction} \label{sec:Introduction}

Molecular crystals are a class of solid composed of individual
molecules bound together by relatively weak intermolecular
interactions, such as van der Waals
forces\cite{silinsh_organic_1994}. Molecular solids have been the
subject of extensive interest over the years, thanks to their actual
or potential applications in fields such as organic
LEDs\cite{zou_recent_2020}, ferroelectrics\cite{khatun_review_2019},
pharmaceuticals\cite{duggirala_pharmaceutical_2015}, and solar
cells\cite{zhou_all-small-molecule_2019}. One of the key features of
many molecular solids is polymorphism: the ability of the same
compound to form several different structures, depending on the
temperature, pressure, and other conditions. Importantly, this
difference in structure can manifest itself in differences between the
observable properties of the polymorphs; for example, different
polymorphs of pharmaceutical compounds can vary in their
pharmaceutical effectiveness, creating both challenges and
opportunities for drug development\cite{censi_polymorph_2015}.

Among the most important properties of a molecular crystal,
particularly for solar cells or similar applications, is its optical
absorption. When different polymorphs of the same compound exhibit
different optical absorption properties, this is known as colour
polymorphism\cite{nogueira_color_2020}. Such compounds have several
potential uses\cite{lu_recent_2019}, including temperature or
mechanical
sensors\cite{gentili_time-temperature_2013,li_mechanochromism_2016},
lasers\cite{liao_cluster-mediated_2018}, and organic field effect
transistors\cite{hsieh_polymorphic_2018}. If the optical absorption
properties of the polymorphs of a candidate compound can be predicted
ahead of time, this can help direct experimental efforts towards more
promising compounds, saving resources and accelerating discovery. One
way these predictions can be made is by using first-principles
materials modelling methods to describe the electronic structure, such
as density functional theory (DFT). Extensive research efforts have
been focused on predicting the crystal structures of polymorphs of
compounds\cite{price_control_2018,beran_modeling_2016} using
high-throughput workflows, allowing the structure of polymorphs to be
predicted from first principles accurately and at reasonable
computational cost. The question is then: can we do something similar
for the optical absorption spectra of these compounds?

One of the most popular methods for calculating absorption spectra of
molecular crystals is time-dependent density functional theory
(TDDFT), thanks to its balance of accuracy and computational
efficiency\cite{kronik_excited-state_2016,zuehlsdorff_linear-scaling_2013}
compared to other methods, such as many-body perturbation theory via
the GW-BSE approach\cite{ulpe_gw-bse_2020}. Previous work on molecular
crystals using TDDFT has largely focused on improving accuracy by
using more complex density
functionals\cite{kronik_excited-state_2016,fonari_impact_2014,bisti_fingerprints_2011,leng_excitons_2015,luftner_experimental_2014},
such as optimally-tuned screened range-separated hybrid (OT-SRSH)
functionals\cite{manna_quantitative_2018,bhandari_fundamental_2018},
but such calculations can be prohibitively computationally expensive
to perform routinely, especially as the system size increases.

A computationally efficient alternative is to calculate the optical
properties of an extended system using computationally efficient, but
less accurate, semi-local TDDFT, and to apply a suitable correction to
the spectrum. Previous work has shown that the unphysical tendency of
semi-local TDDFT to over-delocalise excitations can be counteracted by
constraining the excitations to be localised in real
space\cite{zuehlsdorff_linear-scaling_2013,zuehlsdorff_solvent_2016,zuehlsdorff_predicting_2017}. This
in turn necessitates the use of supercells to converge the real-space
environment in which the excitation is embedded. As for the correction
to the spectrum, this problem is similar to that faced in the
calculation of optical absorption spectra of chromophores in solvent:
long-range interactions between the chromophore and solvent can very
strongly affect the absorption
spectrum\cite{zuehlsdorff_solvent_2016}. To address this problem for
solvated chromophores, Ge \textit{et al.}  introduced a method in
which semi-local TDDFT calculations of a large system consisting of a
chromophore in explicit solvent are corrected using hybrid TDDFT
calculations on isolated chromophores (or chromophores interacting
with only a few solvent molecules)\cite{ge_accurate_2015}. This
so-called \textit{spectral warping} method\footnote[3]{Also known as
  \textit{colour morphing}. Note that the spectral warping method
  referred to here should not be confused with the time domain
  transformation often used in electronics.} has since been used
successfully for a range of chromophores in different
solvents\cite{zuehlsdorff_predicting_2017}.

In this paper, we present a method, based on the spectral warping
method of Ge \textit{et al.},\cite{ge_accurate_2015} for accurately
calculating the optical absorption properties of molecular crystals,
in particular colour polymorphs. Semi-local TDDFT calculations are
performed on a large periodic supercell (referred to as `semi-local
crystal' calculations from here on), and a correction to the
semi-local TDDFT results is then applied, derived from hybrid TDDFT
calculations on dimers (referred to as `hybrid dimer' calculations
from here on). In this work, we refer to this method as the crystal
spectral warping method. We demonstrate the power of the crystal
spectral warping method by applying it to three polymorphs of the
strongly colour polymorphic compound ROY,\cite{yu_polymorphism_2010}
and show that it significantly outperforms semi-local crystal and
hybrid dimer calculations in terms of predicting experimental
spectra. Our crystal spectral warping approach also gives better
agreement with experiment than TDDFT calculations on molecular crystal
supercells using the PBE0 hybrid functional, whilst having
significantly lower computational cost.

\section{The compound ROY} \label{sec:ROY}

\begin{figure}[h!]
  \centering
  \includegraphics[trim={5cm 13.7cm 7cm 5.4cm},clip,width=0.48\textwidth]{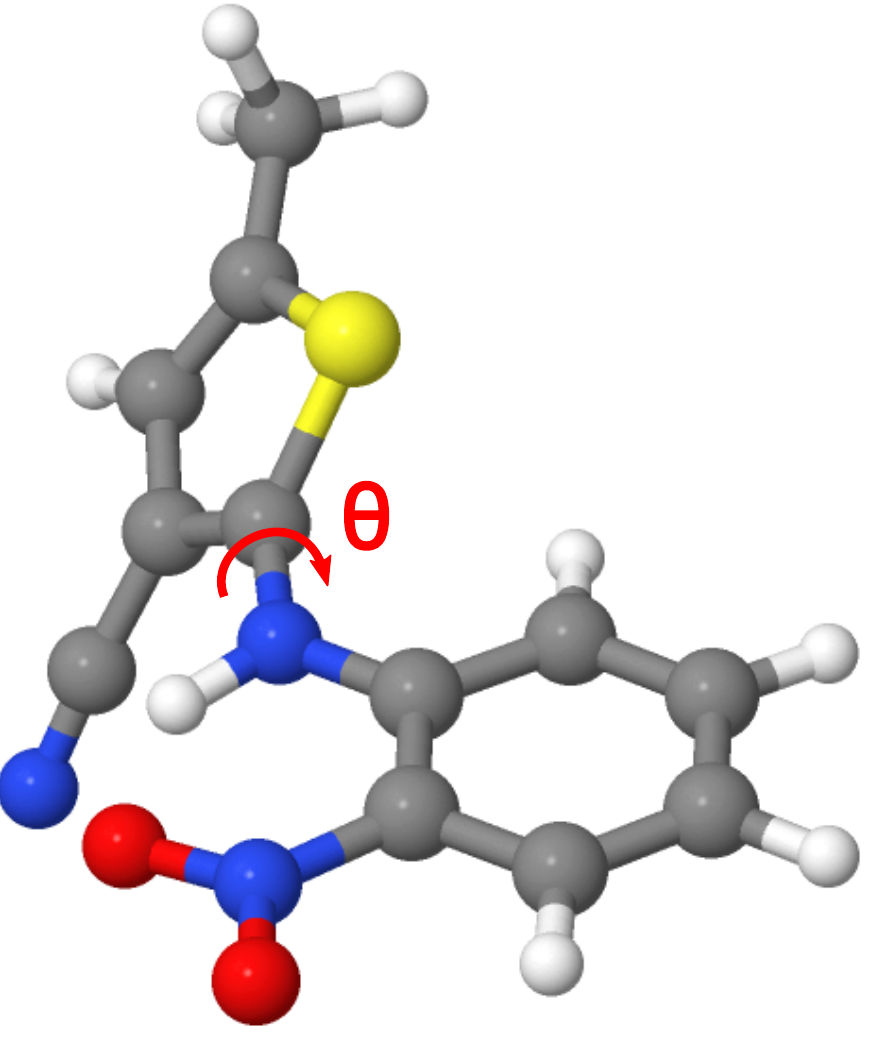}
  \caption{The molecular structure of the ROY molecule. H, C, N, O,
    and S atoms are white, grey, blue, red, and yellow
    respectively. Figure made with Jmol\cite{jmol}. Also labelled is
    the most important dihedral angle $\theta$, which controls how
    planar the molecule is, and is strongly correlated with the colour
    of the molecule.}
  \label{fig:ROYBasics}
\end{figure}

One of the most famous examples of colour polymorphism is
5-methyl-2-[(2-nitrophenyl)amino]-3-thiophenecarbonitrile, or ROY for
short (see Fig.\ \ref{fig:ROYBasics}). When first described in 1995,
five polymorphs were known, which displayed red, orange and yellow
colours (hence the name ROY)\cite{stephenson_conformational_1995}. By
2010, another five polymorphs had been discovered
\cite{yu_thermochemistry_2000,mitchell_selective_2001,chen_new_2005,chen_cross-nucleation_2005},
making it one of the most polymorphic compounds known up to that
point\cite{yu_polymorphism_2010}. Another polymorph was reported in
2019\cite{gushurst_po13_2019}, bringing the number to eleven: nine
whose crystal structures have been found using X-ray diffraction (R,
Y, ON, OP, YN, ORP, YT04, R05, and PO13), and two whose structures
remain unknown (RPL and YT04)\cite{tan_roy_2018, nogueira_color_2020}.

ROY's polymorphs exhibit differences in both molecular conformation,
and in crystal packing. Each polymorph is typically named after its
colour, and, if necessary, a distinguishing characteristic (e.g., N
for needle, P for plate) or the year it was found. The Y polymorph
exhibits piezochromism\cite{harty_reversible_2015}.

It is known experimentally that the most important predictor of the
colour of a polymorph of ROY is the angle $\theta$, which describes
how planar the molecule is (see Fig.\ \ref{fig:ROYBasics}). Values of
$\theta$ between approximately $20^\circ$ and $45^\circ$ correspond to
red polymorphs, between $45^\circ$ and $60^\circ$ to orange
polymorphs, and $60^\circ$ to $80^\circ$ to yellow
polymorphs\cite{nogueira_color_2020}. Lower values of $\theta$ lead to
more conjugation within the molecule, which affects the electronic
states -- most importantly, the states near the band edges -- and thus
the absorption spectrum. Computational work on ROY has mostly focused
on establishing the relative stability of the
polymorphs\cite{nogueira_color_2020,vasileiadis_polymorphs_2012},
rather than the absorption spectra. The only published
first-principles calculations of the optical absorption properties of
ROY (that we are aware of) considered single molecules extracted from
the various polymorphs\cite{yu_color_2002}; the absorption spectra
were calculated using several different methods, including TDDFT, and
an empirical red-shift to account for the crystalline environment was
applied, estimated from experimental data. In the present study, we go
substantially beyond this, by not just considering the absorption due
to isolated molecules, and by calculating the crystal red-shift from
first principles.

We focus on three of the polymorphs of ROY -- R, ON, and YN, between
them representing the full range of colours present in ROY -- as test
cases for our method.

\section{Methods} \label{sec:Methods}

\subsection{Structures of the polymorphs} \label{subsec:Structures}

\begin{figure*}[h!]
  \centering
  \subcaptionbox{R}{
    \includegraphics[trim={1cm 9cm 1cm 9cm},clip,width=0.3\textwidth]{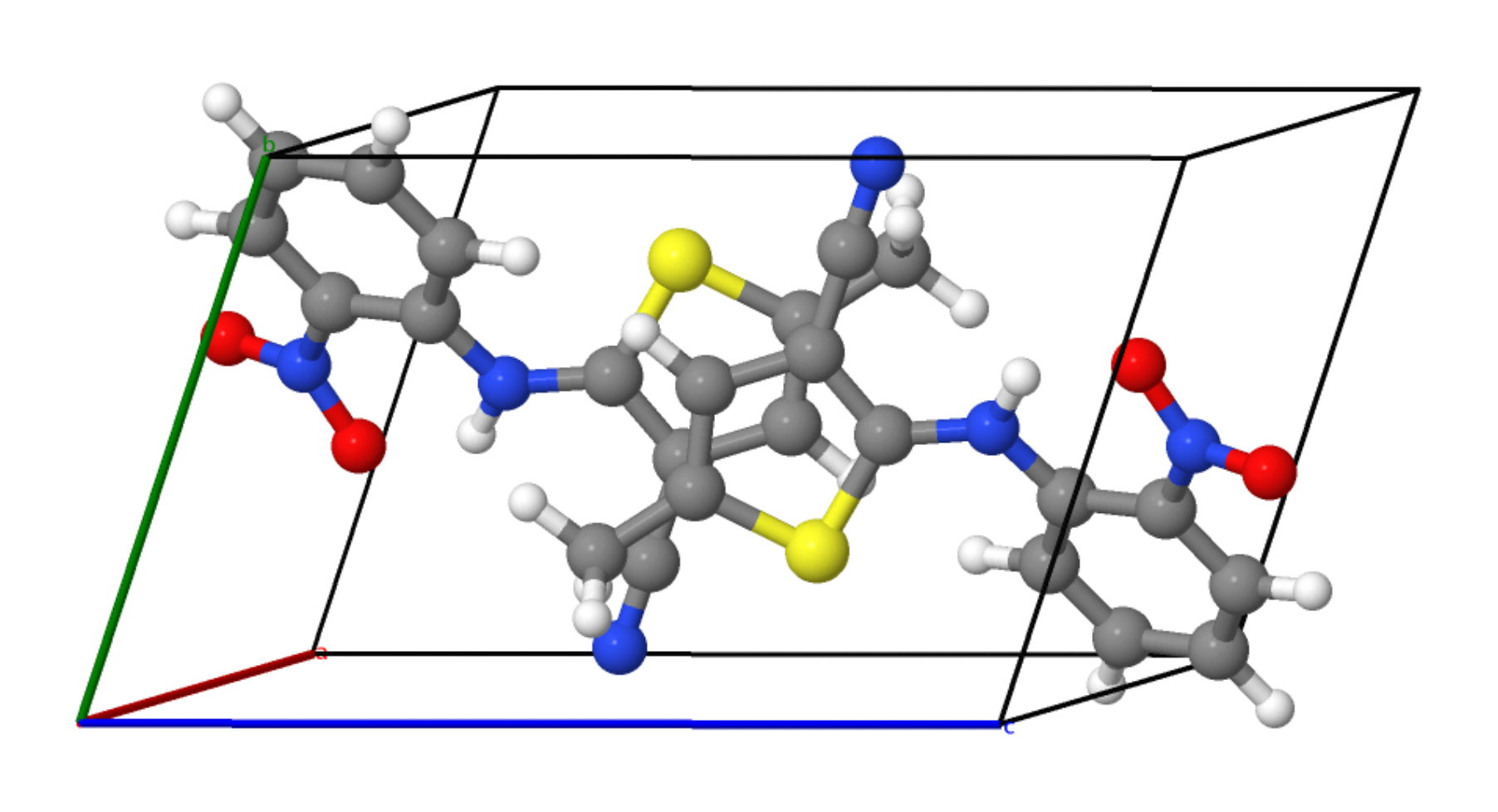}
  }
  ~
  \subcaptionbox{ON}{
    \includegraphics[trim={3cm 6.5cm 3cm 6.5cm},clip,width=0.3\textwidth]{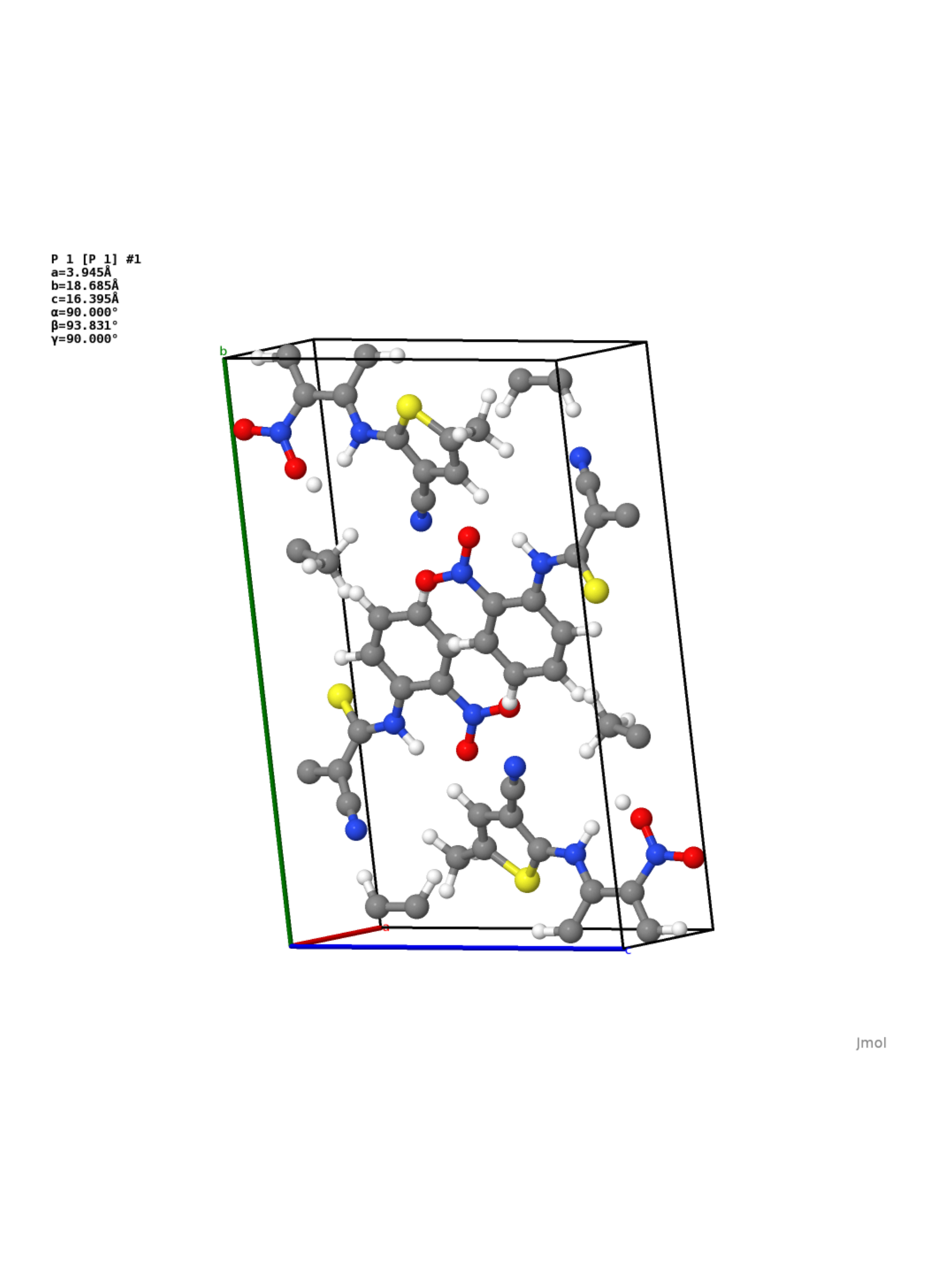}
  }
  ~
  \subcaptionbox{YN}{
    \includegraphics[trim={1cm 8cm 1cm 8cm},clip,width=0.3\textwidth]{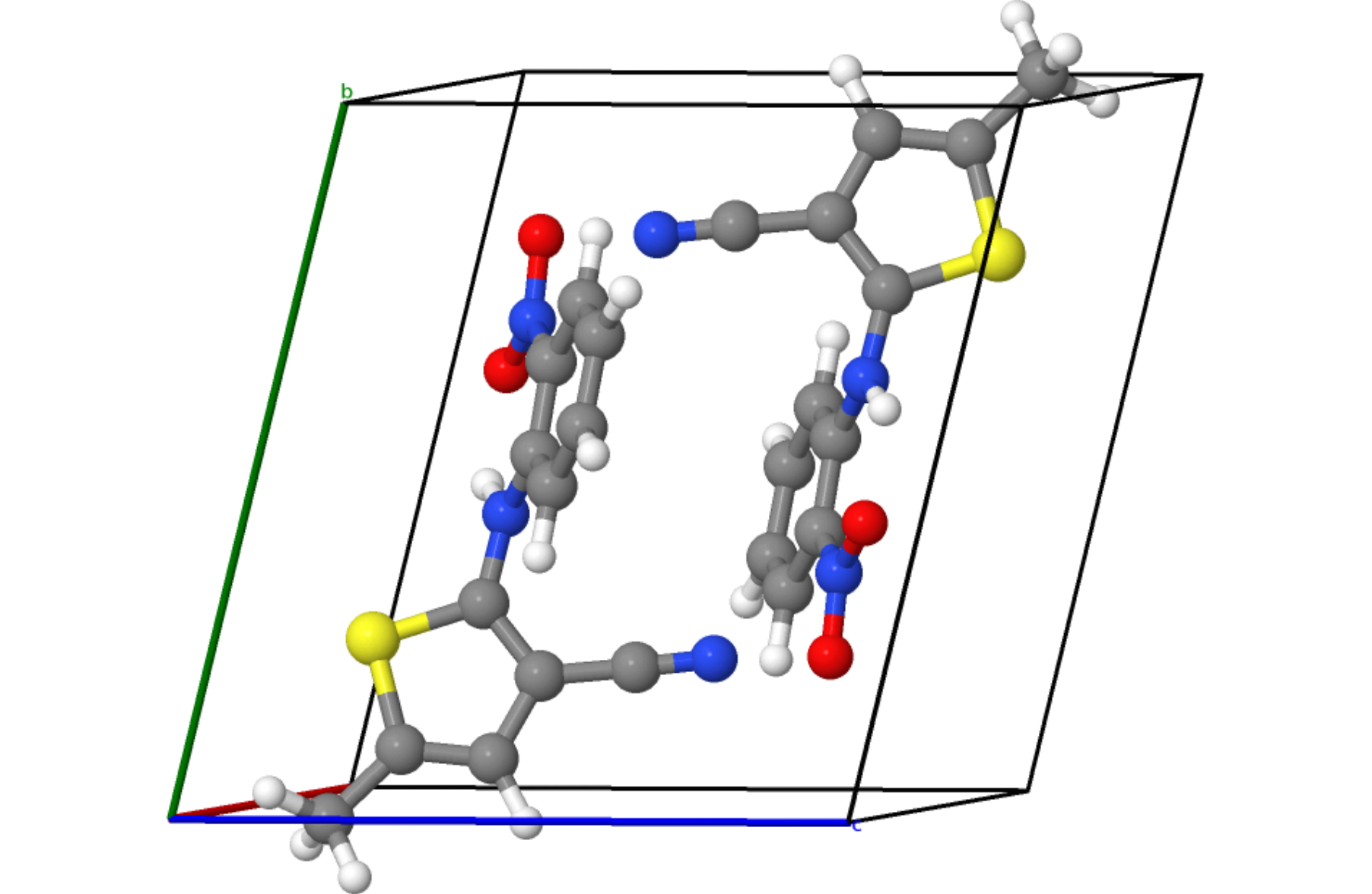}
  }
  \caption{The crystalline unit cells of the three polymorphs of ROY
    considered in this work. The $a$, $b$, and $c$ axes are coloured
    red, green, and blue respectively. H, C, N, O, and S atoms are
    white, grey, blue, red, and yellow respectively. Figures produced
    using Jmol\cite{jmol}.}
  \label{fig:PolymorphStructures}
\end{figure*}

Three polymorphs of ROY are considered in this work: R, ON, and YN. YN
and R both have $P\bar{1}$ symmetry and two molecular units in the unit
cell, whilst ON has a higher $P2_1$ symmetry and four molecular units in
the unit cell. The structures are shown in
Fig.\ \ref{fig:PolymorphStructures}. In the ON polymorph, each pair of
nearest neighbour dimers is essentially equivalent, unlike in the YN
and R polymorphs, where there are several different inequivalent
nearest neighbour dimers -- YN has two, and R has four. 

Experimental structures were obtained from the Cambridge Structural
Database\cite{groom_cambridge_2016}, with the data originally sourced
from Ref.\ \citenum{yu_polymorphism_2010}. These experimental
structures were all measured at or near room temperature. Experimental
structural data from X-ray spectroscopy will typically contain some
uncertainties, particularly in the positions of light elements; in
contrast, however, larger-scale structural features, such as the
lattice parameters or the important dihedral angle $\theta$, will be
more accurately measured by such experiments. In order to reduce these
errors, we optimise the experimental geometry using semi-local DFT,
with lattice constants fixed at the experimental values. Fixing
$\theta$ during the optimisation constitutes a non-linear constraint
which is difficult to impose in practice -- instead, for each
polymorph we use an optimised geometry with as accurate a value of
$\theta$ as possible, as discussed below.

Standard semi-local DFT does not include dispersion interactions, such
as van der Waals forces, although these can be included easily using
various semi-empirical schemes. Including dispersion interactions most
strongly affects the lattice constants, through their effect on
intermolecular binding energies. We can therefore implicitly include a
large portion of the effect of dispersion interactions by fixing the
lattice constants at their room temperature experimental values. This
also facilitates comparison of our results with experimental
measurements taken at room temperature. Using a dispersion correction
scheme in our calculations should then provide small adjustments to
our results, although since all schemes are approximate, these
adjustments will come with errors too. These adjustments could lead to
an optimised value of $\theta$ closer to the experimental value than
by optimising the structure with semi-local DFT alone. As we cannot
fix $\theta$ at its experimental value, we instead compare this to the
value of $\theta$ obtained by relaxing the structures with and without
the explicit inclusion of dispersion interactions, and select the
structure with the value closest to experiment. This allows us to have
a structure where the worst experimental errors have been relaxed
away, but the important larger-scale parameters ($\theta$ and the
lattice parameters) are held at or close to their experimental values.

Here, dispersion interactions are explicitly included using the
Tkatchenko-Scheffler (TS) scheme\cite{tkatchenko_accurate_2009}. The
results of the relaxations (with experimental lattice constants fixed)
both with and without the TS scheme can be found in Section S2 of the
supporting information. We find that for the R and YN polymorphs,
semi-local DFT alone gives the closest match to experiment, whilst for
the ON polymorph, including dispersion interactions gives a better
match to experiment. Because of this, we use the structure optimised
with semi-local DFT only for the R and YN polymorphs, and the
structure optimised with semi-local DFT + TS for the ON
polymorph. These relaxed structures are the ones shown in
Fig.\ \ref{fig:PolymorphStructures}.

\subsection{Calculation of absorption spectra} \label{subsec:SpectraCalc}

As previously discussed, the crystal spectral warping method used in
this work is an extension of the spectral warping method for
calculating the absorption spectra of molecules in
solvent\cite{ge_accurate_2015}. The spectral warping method is
motivated by the fact that a hybrid functional is usually necessary to
obtain quantitatively accurate optical absorption spectra from TDDFT
calculations, but such calculations are often prohibitively
computationally costly for large systems, such as molecules in
explicit solvent, or supercells of a molecular crystal. Using a
semi-local functional is much less computationally expensive for these
systems, but is generally much less accurate. Nonetheless, the shape
of the spectrum at low energies produced by a semi-local TDDFT
calculation is often qualitatively similar to that obtained using a
more accurate hybrid functional. The main differences are often a
shift in the energies of the transitions and a rescaling of oscillator
strengths\cite{zuehlsdorff_solvent_2016,ge_accurate_2015,malcioglu_dielectric_2011}. This
implies that we should be able to apply a (series of) linear
transformation(s) to the semi-local TDDFT spectrum, in order to obtain
a good estimate of the hybrid TDDFT spectrum. This is known as
spectral warping. For each excitation of interest, we can calculate
the estimated hybrid excitation energy $\omega_{\text{hybrid}}$ from
the semi-local excitation energy $\omega_{\text{semi-local}}$ via the
transformation
\begin{equation}
  \omega_{\text{hybrid}} = \alpha \omega_{\text{semi-local}} + \beta ~. \label{eq:SpecWarpEq}
\end{equation}
The oscillator strength $f$ is transformed in a similar way:
\begin{equation}
  f_{\text{hybrid}} = sf_{\text{semi-local}} ~. \label{eq:SpecWarpEq2}
\end{equation}
This observation is unhelpful, however, unless the parameters
$\alpha$, $\beta$, and $s$ can be obtained without performing a
large-scale (supercell) hybrid TDDFT calculation. The question is now
how to obtain these parameters without performing such a calculation,
so that the transformations in Equations \eqref{eq:SpecWarpEq} and
\eqref{eq:SpecWarpEq2} can be used to produce an approximate
`large-scale hybrid TDDFT' spectrum from a large-scale semi-local
TDDFT spectrum.

\begin{figure}[h!]
  \centering
  \includegraphics[trim={2cm 0cm 1cm 2cm},clip,width=\textwidth]{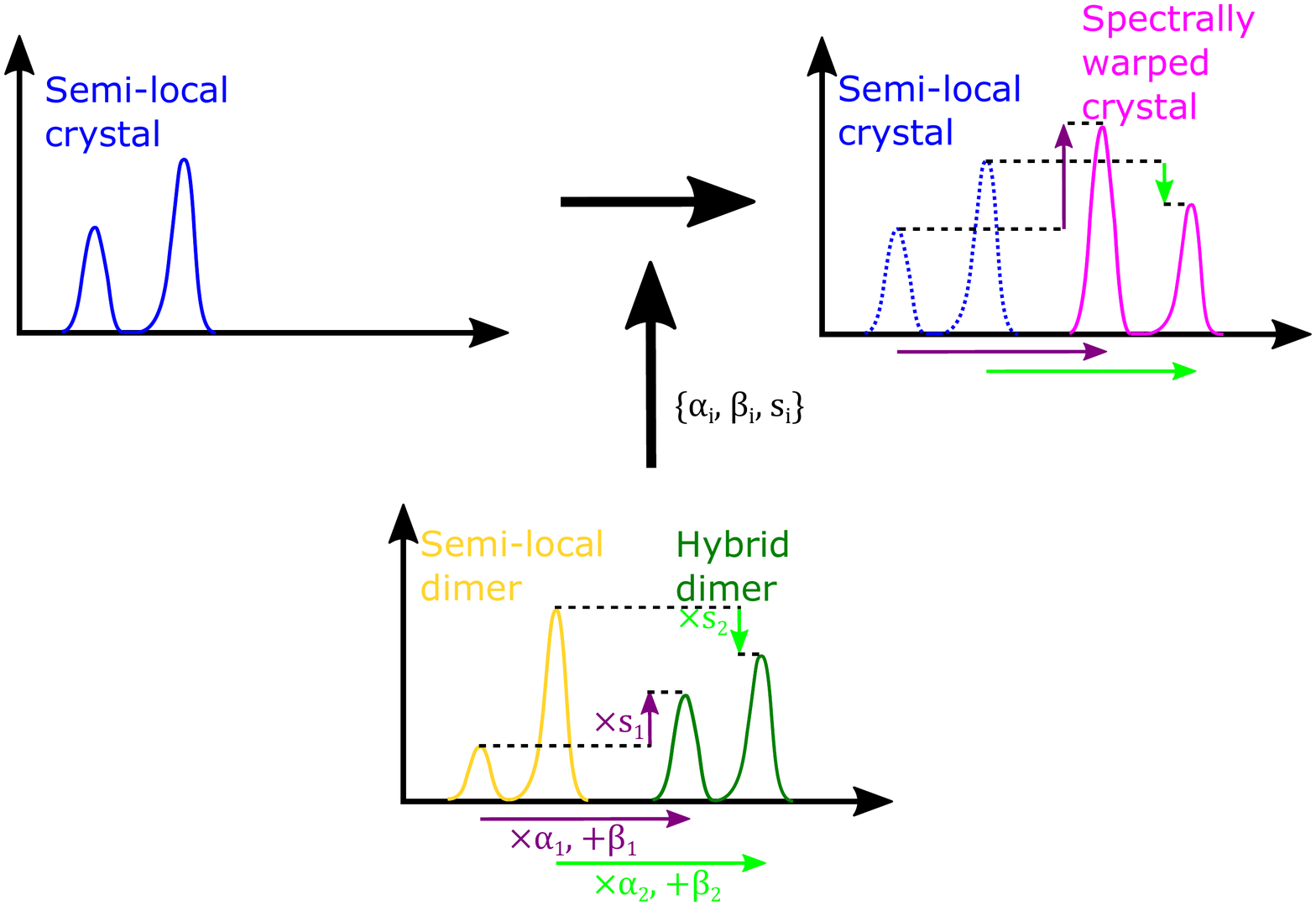}
  \caption{Schematic showing the crystal spectral warping process. By
    comparing semi-local dimer calculations (gold curve) and hybrid
    dimer calculations (dark green curve), we can determine linear
    transformations for the frequency and oscillator strength of each
    excitation (purple and light green arrows). We can then apply
    these transformations to semi-local crystal calculations (blue
    curve), producing a spectrally warped spectrum (magenta curve).}
  \label{fig:SpecWarpSchematic}
\end{figure}

Previous work on molecules in solvent has determined these parameters
by making use of the fact that the excitations of interest are largely
localised on a particular part of the system, i.e., the solute. This
sub-system is small enough that we can perform semi-local \textit{and}
hybrid TDDFT calculations on it in isolation. By mapping the former
calculation onto the latter, we can obtain values for $\alpha$,
$\beta$ and $s$ for each relevant excitation. This mapping process is
performed manually, by comparing the main orbitals involved in the
excitations. These values for $\alpha$, $\beta$ and $s$ can then be
inserted into Equations \eqref{eq:SpecWarpEq} and
\eqref{eq:SpecWarpEq2}, which can then be used to separately warp each
excitation of the larger system calculated using semi-local TDDFT,
generating an absorption spectrum that approximates the hybrid TDDFT
absorption spectrum of the larger system. This process, as used in
this work, is shown schematically in
Fig.\ \ref{fig:SpecWarpSchematic}. The advantage of this technique is
that it includes the effects of the wider environment on the
excitations, as well as treating the excitations accurately with
hybrid functionals. We might expect that the absorption spectrum of a
molecular crystal such as ROY will be dominated by effects arising
from interactions between nearest neighbour molecules (i.e., molecular
dimers), with the influence of more distant molecules providing a
smaller screening correction. Calculations on molecular clusters of
increasing size -- presented in Section S7.2 of the supporting
information -- support this hypothesis. This suggests that the
spectral warping method may work well for molecular crystals if we use
the results for an isolated dimer to determine the warping parameters
$\alpha$, $\beta$, and $s$. This dimer is `cut out' of the molecular
crystal, and used without any further geometry optimisation once it is
isolated.

We compare the optical absorption spectra obtained by the crystal
spectral warping method to the spectra obtained through the semi-local
crystal and hybrid dimer calculations outlined above. We also compare
our results to experimental measurements. For the YN polymorph we also
compare the results of the crystal spectral warping method to the
results arising from applying a global hybrid functional (PBE0)
directly to a supercell of the molecule crystal. Our reasons for using
a global hybrid here are discussed in Section
\ref{subsec:CompDetails}. A comparison to supercell TDDFT calculations
using a non-optimally tuned range-separated hybrid functional are also
presented in Section S3 of the supporting information.

\subsection{Computational details} \label{subsec:CompDetails}

The geometry optimisations discussed in Section
\ref{subsec:Structures} were performed with the plane-wave
pseudopotential DFT code CASTEP\cite{clark_first_2005} using the PBE
functional\cite{perdew_generalized_1996}. As previously noted,
dispersion interactions are included using the TS
scheme\cite{tkatchenko_accurate_2009} for the ON polymorph only. These
calculations were all run in periodic boundary conditions using only
the unit cell of the polymorph in question. CASTEP's on-the-fly (OTF)
norm-conserving pseudopotentials were used; the corresponding
pseudopotential strings can be found in Section S1 of the supporting
information. We used a kinetic energy cut-off of $1000$~eV, and
electronic k-point sampling was performed in CASTEP using a
Monkhorst-Pack grid\cite{monkhorst_special_1976} with a spacing of
$0.03$~\r{A}$^{-1}$.

The semi-local crystal TDDFT calculations, which also act as the base
for the crystal spectral warping method, were performed with the code
ONETEP\cite{prentice_onetep_2020}, using the PBE
functional\cite{perdew_generalized_1996}. As semi-local TDDFT has a
well-known issue with over-delocalisation, which leads to incorrect
results, we enforce localisation by restricting the excitations to be
localised on a given dimer, by truncating the response density kernel
appropriately, similar to the approach used previous
work\cite{zuehlsdorff_solvent_2016}. This is possible because ONETEP
uses a set of localised non-orthogonal generalised Wannier functions
(NGWFs) to represent the electronic structure. How appropriate dimers
are selected is discussed in more detail below. The real-space
constraint on the excitations means that a supercell must be used to
converge the size of the real-space environment in which the
excitation is embedded. We used a supercell consisting of
$4\times3\times2$ unit cells (containing $48$ molecular units) for the
YN and R polymorphs, whilst a $4\times1\times2$ supercell (containing
$32$ molecular units) was used for the ON polymorph. Convergence
calculations showed that the supercells used in this work were
sufficiently large to obtain highly accurate results -- for more
details, see Section S7.1 of the supporting information. As a
linear-scaling DFT code, ONETEP allows us to perform these large-scale
calculations both efficiently and accurately. These calculations were
performed in periodic boundary conditions. For consistency, we used
the same pseudopotentials and kinetic energy cut-off for our ONETEP
calculations as those used in our CASTEP calculations. Unlike CASTEP,
ONETEP only samples the $\Gamma$-point, as it is designed for use on
large-scale systems (such as the supercells used here) where this is
sufficient. More details of the calculational parameters used can be
found in Section S1 of the supporting information.

The semi-local and hybrid dimer TDDFT calculations, which are used to
obtain the parameters for the spectral warping transformations within
the crystal spectral warping method, were performed using
NWChem\cite{apra_nwchem_2020} (unless noted), using the PBE and
PBE0\cite{adamo_toward_1999} functionals, respectively. We used PBE0
for this purpose as it has been shown to perform well in previous work
on benchmarking different functionals for TDDFT in aromatic dye
molecules\cite{jacquemin_extensive_2009}, and also because it gives
good results for band gaps in other molecular
systems\cite{fedorov_structure_2017,fedorov_pressure_2019}. PBE0 is
also widely available and well-tested in most TDDFT codes. All NWChem
calculations were performed in open boundary conditions, and the
cc-pVDZ basis set was used for all species. This basis set does not
include diffuse basis functions, which can substantially increase
accuracy for some systems\cite{papajak_perspectives_2011}. However, we
have found that including diffuse basis functions only produces a
small rigid correction to our NWChem results, and that the spectral
warping parameters are barely affected at all. This is due to the
cancellation of systematic errors present in both the PBE and PBE0
calculations, which result in the calculations at different levels of
theory shifting by almost exactly the same amount. As including
diffuse basis functions significantly affects the computational
expense and numerical stability of the
calculations\cite{papajak_perspectives_2011}, we therefore use the
cc-pVDZ basis set. Further discussion of the effects of using diffuse
basis functions can be found in Section S7.3 of the supporting
information.

The hybrid crystal TDDFT calculation for the YN polymorph was
performed with the mixed Gaussian/plane-wave DFT code
CP2K\cite{kuhne_cp2k_2020}, using the hybrid functional PBE0. As
previously noted, these calculations are significantly more expensive
than equivalently-sized semi-local crystal TDDFT calculations,
particularly in terms of computer memory. This limited us to
performing these calculations on only one polymorph (YN), and also to
a smaller supercell than that used elsewhere in the semi-local crystal
TDDFT calculations ($3\times2\times1$ versus $4\times3\times2$). These
calculations were performed in periodic boundary conditions. We used
the cc-TZV2P-GTH basis set for all species except S, for which we used
TZV2P-GTH, and we used the potentials GTH-PBE-q1, -q4, -q6, -q5, and
-q6 for H, C, O, N, and S, respectively. These potentials are those
distributed with CP2K. For hybrid functional calculations in periodic
boundary conditions in CP2K, it is necessary to limit the range of the
exact-exchange part of the functional. For the PBE0 calculations, we
used a truncated Coulomb potential with a cut-off radius of $5.8$
bohr.

To reduce computational cost, the Tamm-Dancoff
approximation\cite{hirata_time-dependent_1999} (TDA) was applied in
our ONETEP and CP2K TDDFT calculations.  This approximation neglects
the coupling between excitation and de-excitation processes, making
the TDDFT eigenvalue problem Hermitian, and thus much easier to
solve. The TDA typically gives accurate excitation frequencies, but
performs worse for the oscillator
strengths\cite{casida_progress_2012,zuehlsdorff_linear-scaling_2013}. In
this work, we are primarily interested in the positions of the peaks
in the absorption spectrum, so the TDA is a reasonable approximation
to take. However, in our NWChem calculations, the TDDFT problem was
solved in full, allowing any error in the oscillator strengths to be
partially corrected by the spectral warping correction. Closer testing
showed that using the TDA rather than full TDDFT for our NWChem
calculations had very little effect on the spectral warping
parameters, thanks to a cancellation of errors similar to that for
diffuse basis functions described above, but we maintain the use of
full TDDFT in our NWChem calculations for the sake of
accuracy. Further discussion of the effects of using the TDA can be
found in Section S7.4 of the supporting information.

One aspect of the calculations that should be particularly noted is
that, as pointed out in Section \ref{subsec:Structures}, there are
several inequivalent nearest neighbour dimers in the YN and R
polymorphs. To build up a full spectrum, therefore, it is necessary to
add together the contributions from each inequivalent dimer. This
means the hybrid dimer results presented in Section \ref{sec:Results}
are actually the sum of the absorption spectra for each of these
isolated dimers. Similarly, the semi-local crystal results presented
are the sum of the absorption spectra with the excitation restricted
to be on each of these dimers in turn. To obtain the crystal spectral
warping results, the spectral warping procedure described above is
applied \textit{to each dimer separately} -- the semi-local TDDFT
calculation with the excitation restricted to a given dimer is warped
using the vacuum results for the same dimer. The warped results for
all dimers are then summed to give the crystal spectral warping
spectra.

Apart from the codes used to perform the actual TDDFT calculations, we
also used c2x\cite{rutter_c2x_2018} to construct supercells, and
Jmol\cite{jmol} and VESTA\cite{momma_vesta_2011} for visualisation
purposes.

\section{Results} \label{sec:Results}

\begin{figure*}[h!]
  \centering
  \subcaptionbox{R}{
    \includegraphics[width=0.47\textwidth]{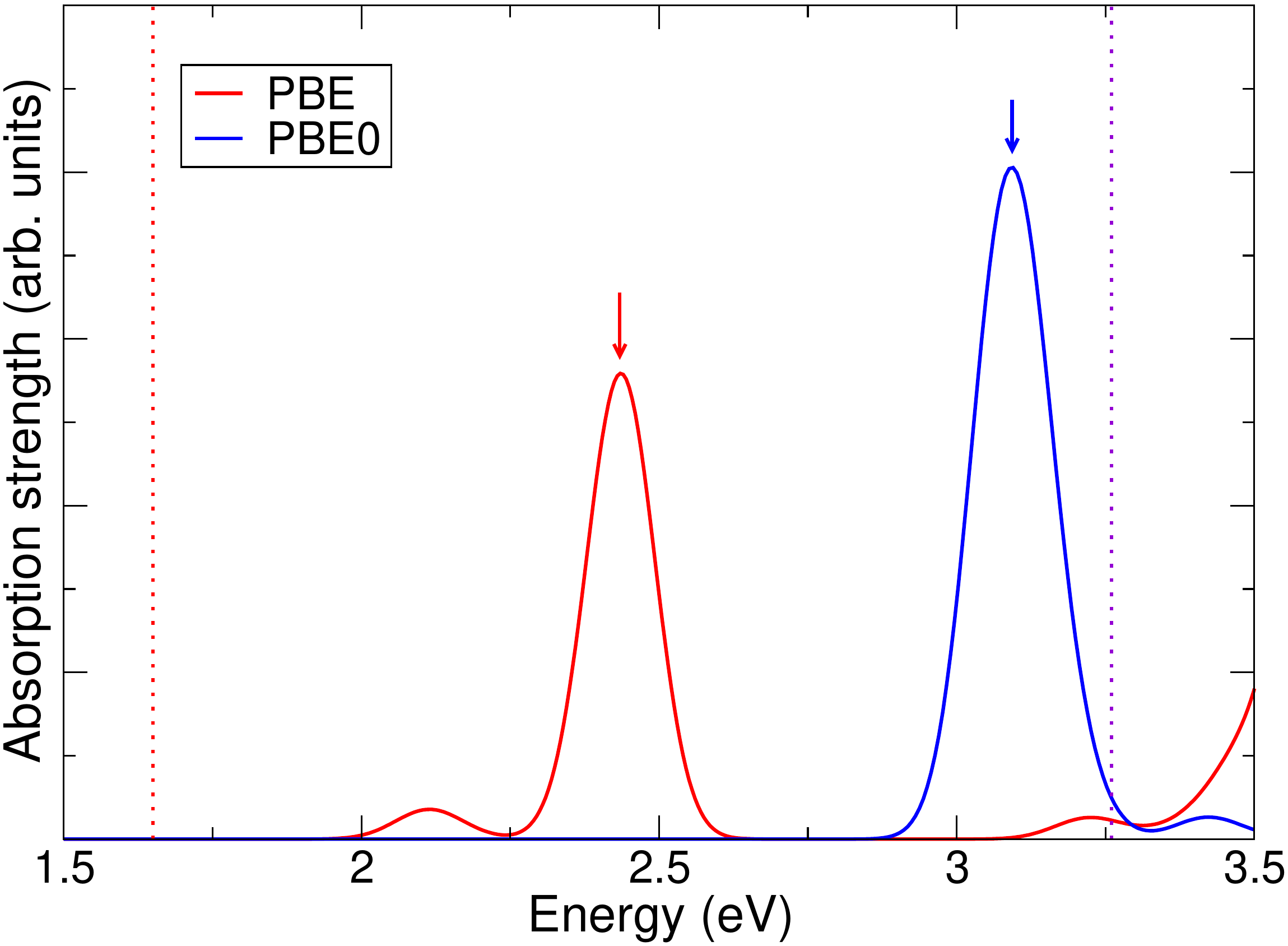}
  }
  ~
  \subcaptionbox{ON}{
    \includegraphics[width=0.47\textwidth]{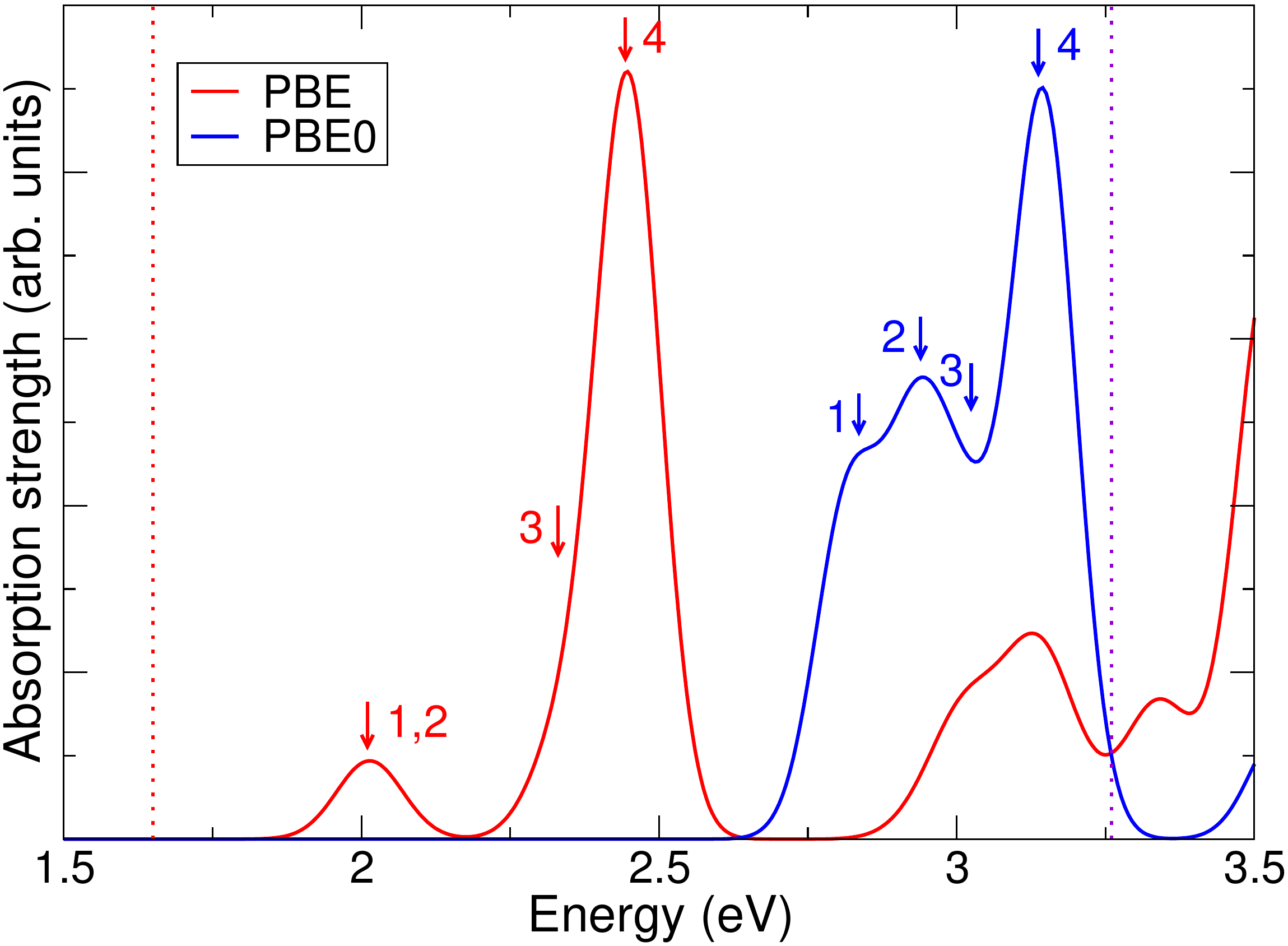}
  }
  ~
  \subcaptionbox{YN}{
    \includegraphics[width=0.47\textwidth]{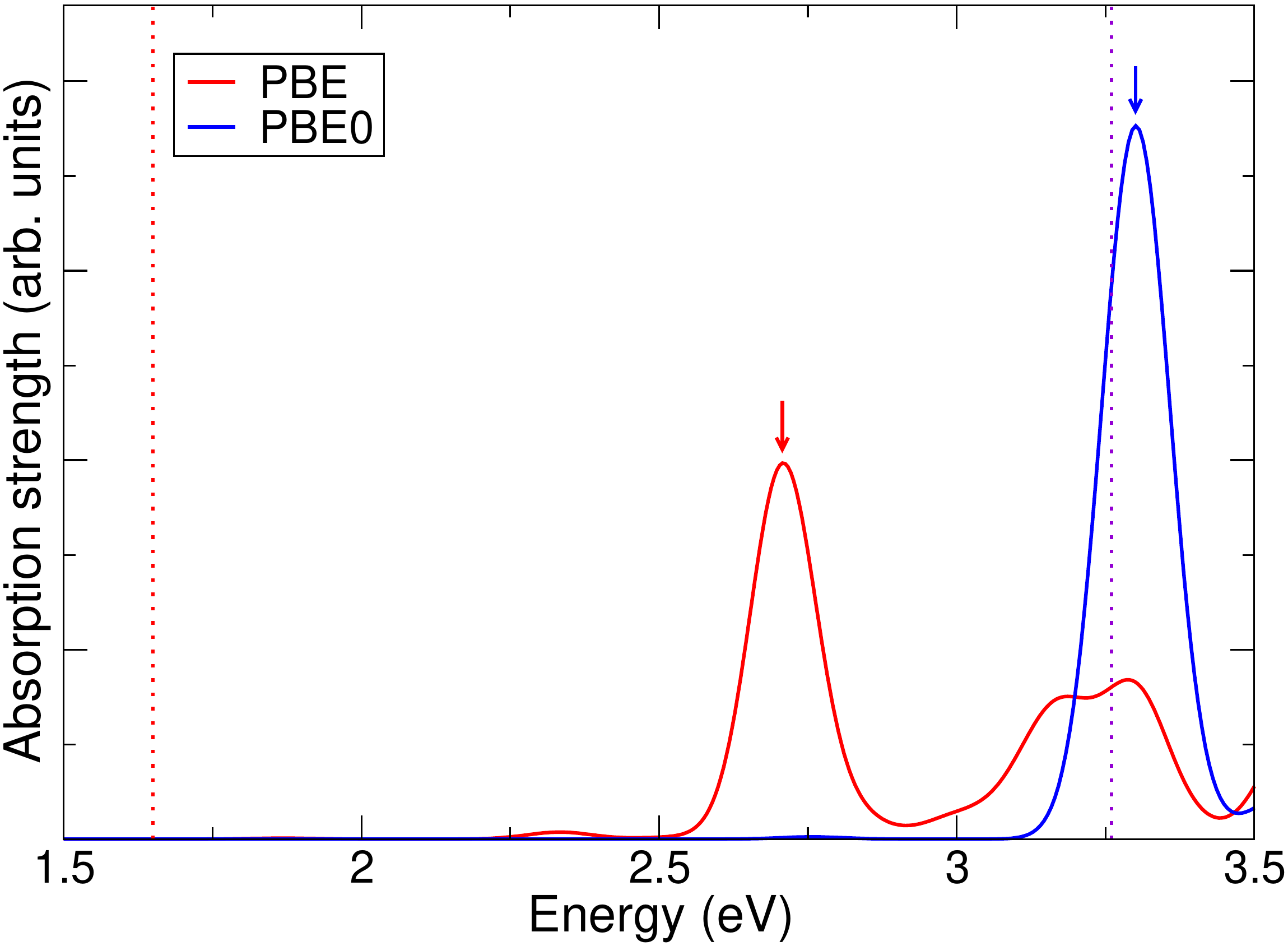}
  }
  \caption{Calculated absorption spectra for isolated dimers extracted
    from the R, ON, and YN polymorphs. Each subfigure shows the
    spectra calculated using both PBE and PBE0 functionals, summed
    over symmetry-inequivalent nearest neighbour dimers from each
    polymorph. These spectra correspond to the sum of the spectra used
    for the spectral warping process. Arrows indicate the peaks used
    to obtain the spectral warping parameters -- in the case of ON,
    corresponding peaks in the PBE and PBE0 spectra are labelled with
    the same number. The PBE0 results correspond to the hybrid dimer
    calculations discussed in the main text. The vertical dotted lines
    show the approximate edges of the visible spectrum.}
  \label{fig:SpecWarpAbsSpectra}
\end{figure*}

\begin{figure*}[h!]
  \centering
  \subcaptionbox{R}{
    \includegraphics[width=0.47\textwidth]{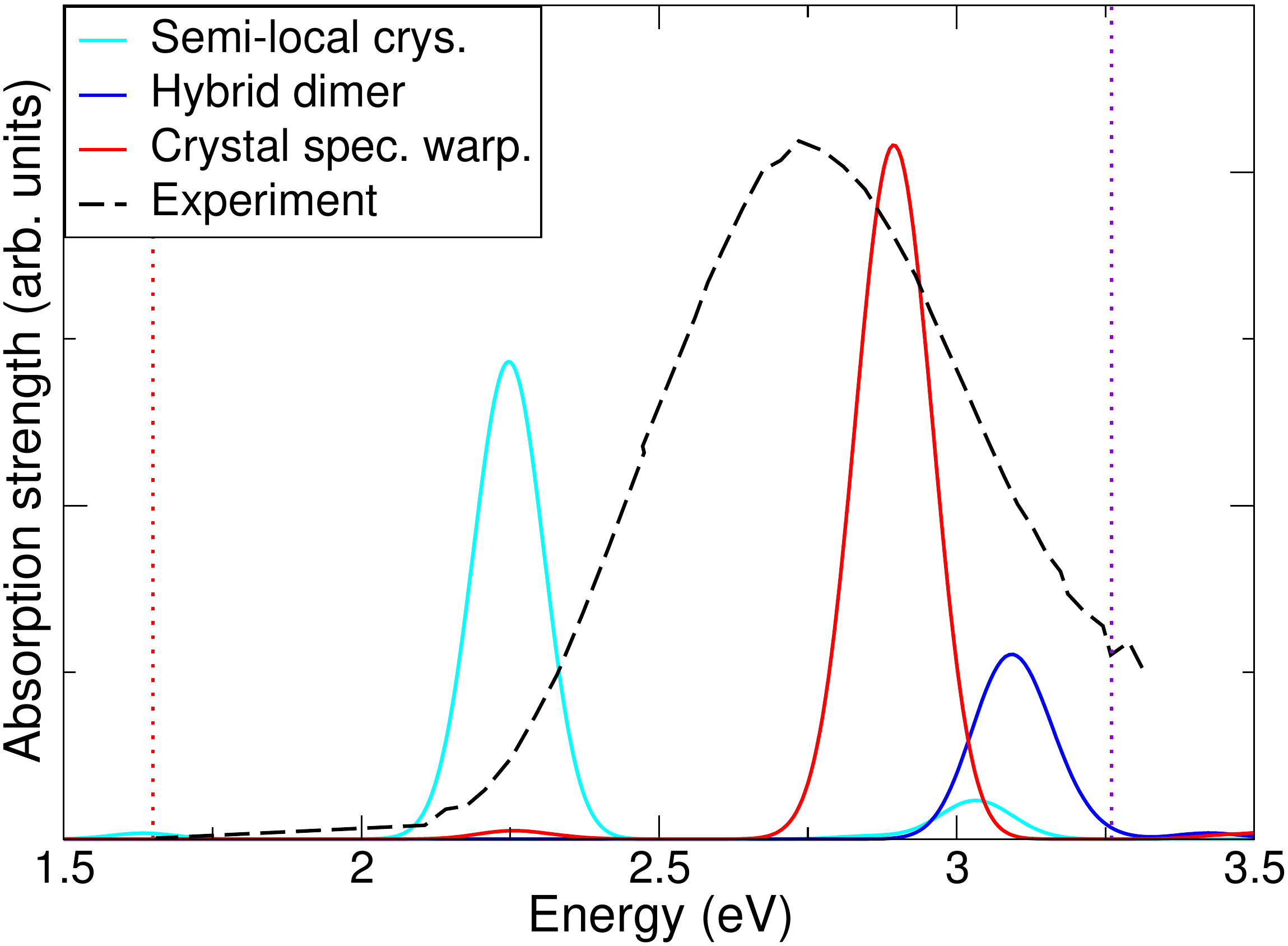}
  }
  ~
  \subcaptionbox{ON}{
    \includegraphics[width=0.47\textwidth]{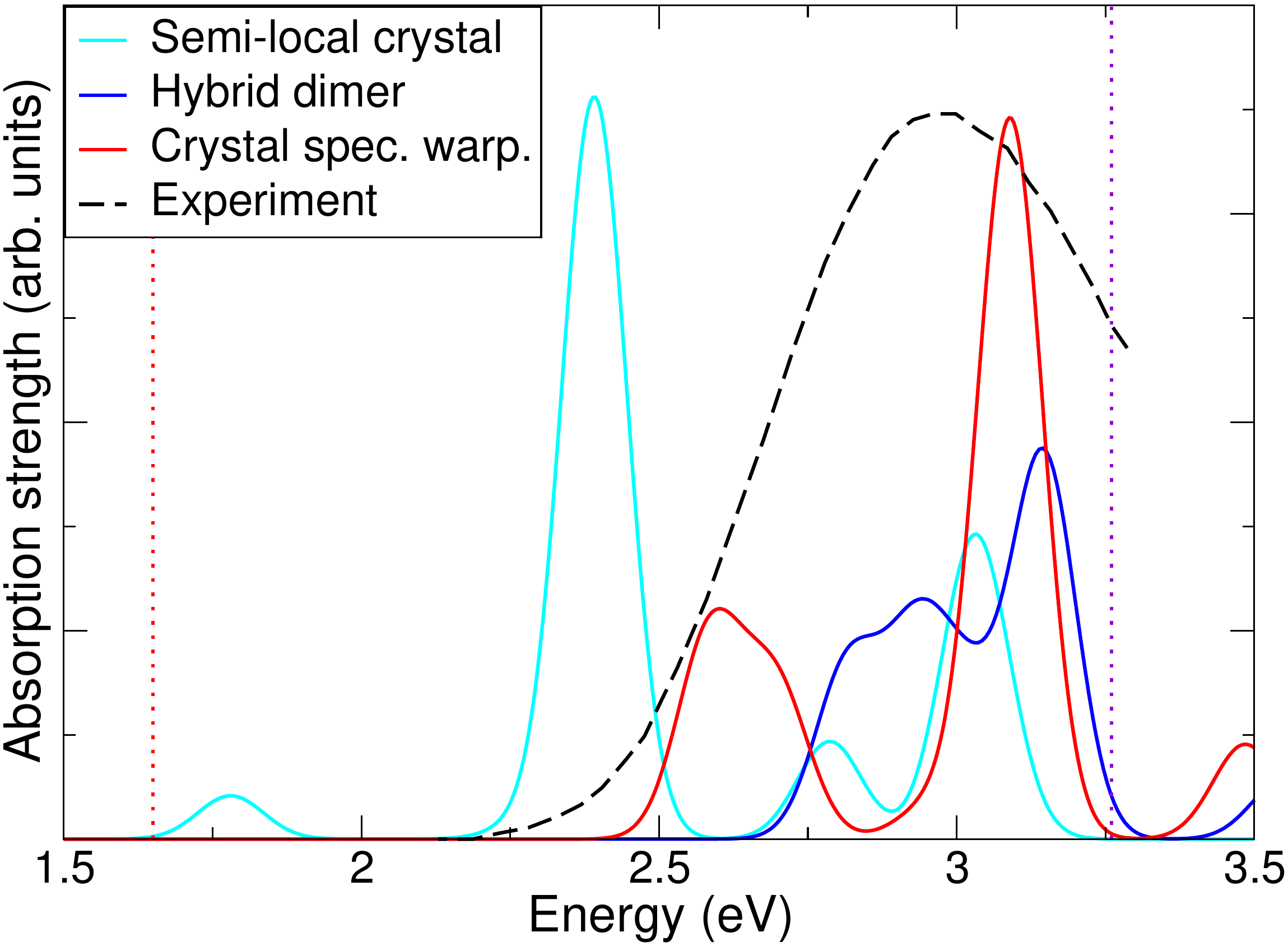}
  }
  ~
  \subcaptionbox{YN}{
    \includegraphics[width=0.47\textwidth]{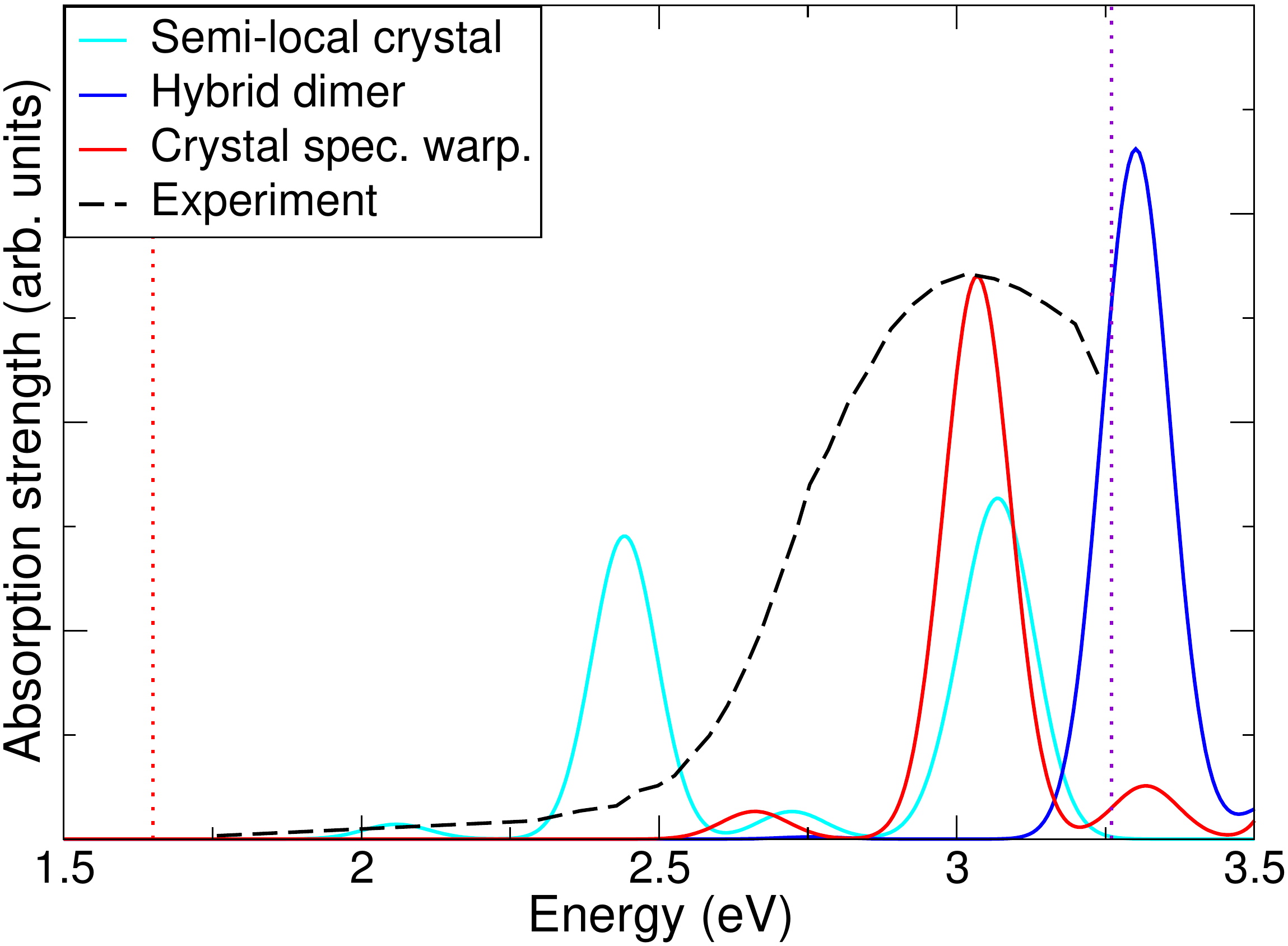}
  }
  \caption{Calculated absorption spectra for the R, ON, and YN
    polymorphs, calculated using several different methods, as
    described in Section \ref{subsec:SpectraCalc}. Each subfigure
    compares the spectra calculated by our crystal spectral warping
    method with the results of semi-local crystal and hybrid dimer
    calculations, as described in the main text, as well as
    experimental data, for each polymorph. The crystal calculations
    were performed using a $4\times3\times2$ supercell for the R and
    YN polymorphs, and a $4\times1\times2$ supercell for the ON
    polymorph. Experimental data is taken from
    Ref.\ \citenum{yu_color_2002}. The vertical dotted lines show the
    approximate edges of the visible spectrum.}
  \label{fig:AbsorptionSpectra}
\end{figure*}

\begin{figure*}[h!]
  \centering
  \includegraphics[trim={5cm 0cm 0cm 1.5cm},clip,width=0.47\textwidth]{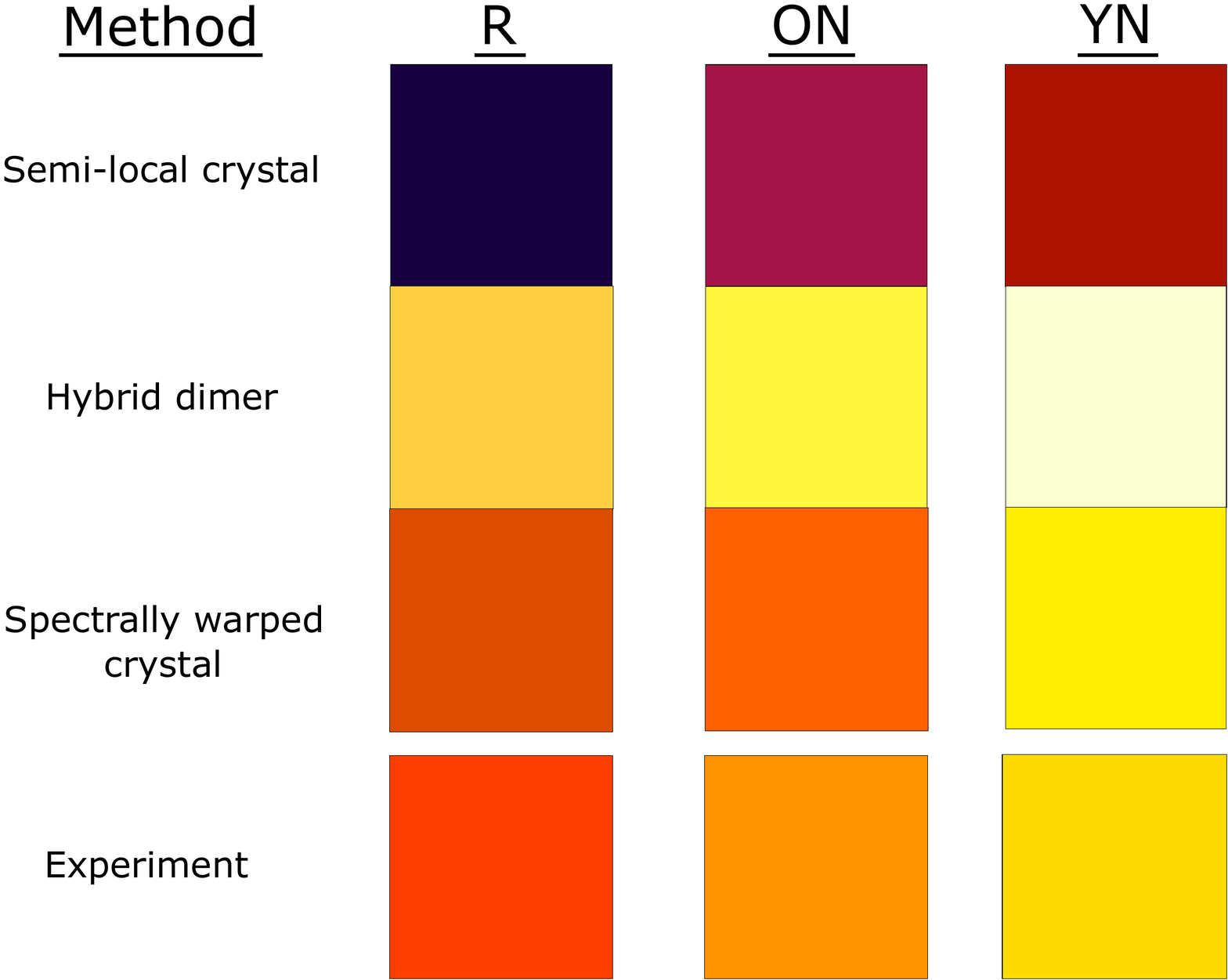}
  \caption{Swatches of colours generated from the absorption spectra
    calculated using several different methods, as well as from
    experimentally measured spectra. The spectra used are those shown
    in Fig.\ \ref{fig:AbsorptionSpectra}. The colour swatches were
    generated using the Colour Python package\cite{colour_package},
    using the method described in Section S4 in the supporting
    information.}
  \label{fig:ColourSwatches}
\end{figure*}

\begin{figure}[h]
  \centering
  \includegraphics[width=0.47\textwidth]{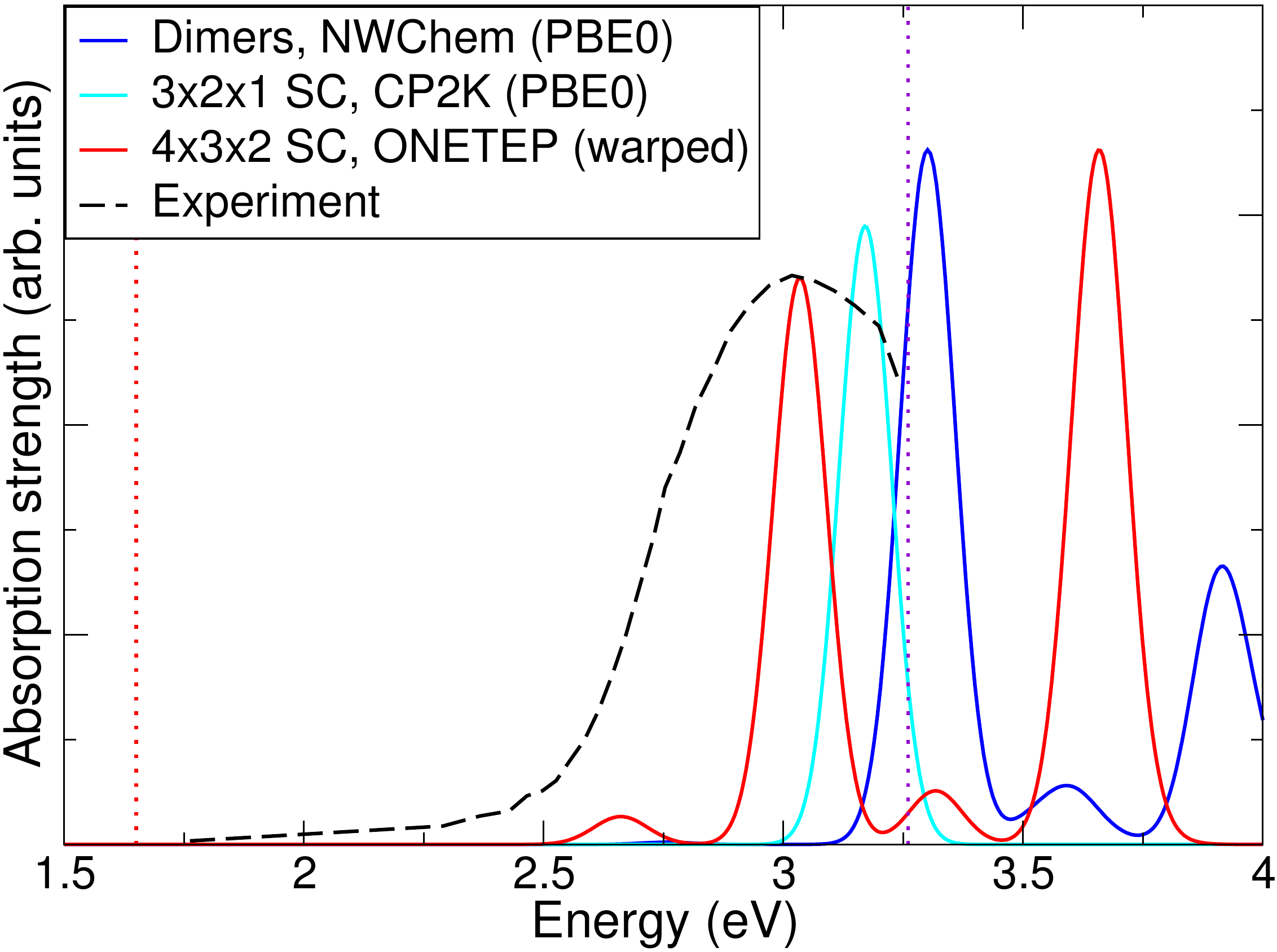}
  \caption{Absorption spectra for the YN polymorph calculated using
    several different methods, as described in Section
    \ref{subsec:SpectraCalc}. Selected results from
    Fig.\ \ref{fig:AbsorptionSpectra}c are compared to the results of
    hybrid TDDFT supercell calculations, performed with the PBE0
    functional. `\textit{$3\times2\times1$ SC, CP2K (PBE0)}' shows the
    spectra calculated with CP2K using PBE0 for a $3\times2\times1$
    supercell. `\textit{$4\times3\times2$ SC, ONETEP (warped)}' and
    `\textit{Dimers, NWChem (PBE0)}' show the same data as
    `\textit{Semi-local crystal}' and `\textit{Hybrid dimer}' in
    Fig.\ \ref{fig:AbsorptionSpectra}c respectively, but have been
    renamed to emphasise the size of supercell and functional
    used. Experimental data is taken from
    Ref.\ \citenum{yu_color_2002}. The vertical dotted lines show the
    approximate edges of the visible spectrum.}
  \label{fig:AbsorptionSpectraRangeSep}
\end{figure}

Before presenting the results of the crystal spectral warping method,
it is important to discuss the spectral warping process
used. Figs.\ \ref{fig:SpecWarpAbsSpectra}a, b, and c show the
absorption spectra used to calculate the spectral warping
transformations for the R, ON, and YN polymorphs respectively, i.e.,
the absorption spectra calculated for isolated dimers using PBE and
PBE0, summed over all symmetry-inequivalent nearest neighbour
dimers. It should be noted that the PBE0 results presented in
Fig.\ \ref{fig:SpecWarpAbsSpectra} correspond to the hybrid dimer
calculations shown in Fig.\ \ref{fig:AbsorptionSpectra}. To obtain
realistic, smoothed spectra, we apply a Gaussian broadening to the raw
TDDFT excitation excitations, with standard deviation of
$0.002$~hartree ($0.0544$~eV). This results in the vast majority of
the broadening shown, as the excitation energies were largely similar
for symmetry-inequivalent dimers. To decide on which excitations
spectral warping should be applied to, we look at the PBE0 results,
and identify any excitations within the visible range that have
significant oscillator strength.

In the R and YN polymorphs, there is only one relevant excitation by
these criteria, making the spectral warping process
straightforward. In the ON polymorph, however, there is more than one
potentially relevant excitation in both the PBE and PBE0 spectra,
meaning we need to perform spectral warping separately on each of
these excitations. In both the PBE and PBE0 spectra, there are 4
peaks. In the PBE spectrum, peaks 1 and 2 have effectively the same
energy, but in the PBE0 spectrum they split, as seen in
Fig.\ \ref{fig:SpecWarpAbsSpectra}b. Peak 4 is by far the strongest in
both the PBE and PBE0 spectra, which means that the nearby peak 3 is
not visible in Fig.\ \ref{fig:SpecWarpAbsSpectra}b, although its
approximate position is labelled. A spectral warping transformation is
found for each corresponding pair of peaks: one mapping the `1' peaks
onto each other, one mapping the `2' peaks onto each other, and so
on. These spectral warping transformations are used to transform the
semi-local crystal calculations for the ON polymorph. In all spectral
warping calculations, we follow previous
work\cite{zuehlsdorff_solvent_2016} and take $\alpha=1$, meaning that
spectral warping effectively corresponds to a rigid frequency shift
and a rescaling of the oscillator strength. The values of $\beta$ and
$s$ resulting from this process are tabulated in Section S5 of the
supporting information.

The absorption spectra calculated using the crystal spectral warping
method for the R, ON, and YN polymorphs are presented in
Figs.\ \ref{fig:AbsorptionSpectra}a, b, and c respectively, and are
compared to hybrid dimer results and experimental
data\cite{yu_color_2002}. The same Gaussian broadening is applied as
in Fig.\ \ref{fig:SpecWarpAbsSpectra}. In the R and YN polymorphs, the
crystal spectral warping method significantly improves on the hybrid
dimer results, predicting a main absorption frequency much closer to
the experimental value. For R, the hybrid dimer calculations predict
an excitation frequency $0.36$~eV above the experimental peak, whilst
the crystal spectral warping method gives a frequency that is only
$0.16$~eV above. For YN, the hybrid dimer calculations predict a
frequency that is $0.28$~eV above the experimental peak, but this
difference is reduced by an order of magnitude by the crystal spectral
warping method, giving a frequency only $0.02$~eV above experiment. In
ON, the picture is slightly more complicated due to the presence of
four absorption peaks, which are likely smeared into one in the
experimental measurement. However, if we look at the main (highest)
peak (corresponding to peak 4 on Fig.\ \ref{fig:SpecWarpAbsSpectra}b),
we can see that the crystal spectral warping method agrees more
closely with experiment than the hybrid dimer calculations. Hybrid
dimer calculations predict an energy of $0.17$~eV away from the
experimental peak for the main peak; using the crystal spectral
warping method, the main peak is $0.11$~eV away. Importantly, the
crystal spectral warping method is closer to experiment than the
hybrid dimer calculations in all polymorphs. More details of how the
crystalline environment affects the character of the excitation are
presented in Section S6 of the supporting information. Interestingly,
these results suggest that, in the full crystalline environment, the
excitations localise even further compared to in an isolated
dimer. This suggests that the potential provided by the surrounding
molecular crystal has the effect of confining the excitation, so that
the excitation in the true molecular crystal is actually almost (but
not completely) monomeric in nature.

The conclusion that the crystal warping method is an improvement over
hybrid dimer calculations is backed up strikingly, although
qualitatively, by generating colour swatches for each method used, as
well as the experimental data. The Colour Python
package\cite{colour_package} is used to do this, following the method
laid out in Section S4 of the supporting
information. Fig.\ \ref{fig:ColourSwatches} shows these swatches for
each polymorph. The swatches corresponding to the experimental data
demonstrate that the three polymorphs are indeed red, orange, and
yellow in colour, as expected. It can be seen that the spectral
warping results give a colour significantly closer than the hybrid
dimer results to the experimental colour. The hybrid dimer and crystal
spectral warping methods give a yellow-orange and brown-red colour
respectively for the R polymorph; for the ON polymorph, the colours
are yellow and red-orange respectively; for the YN polymorph, the
colours are light yellow and yellow respectively. This qualitative
improvement in colour prediction is a direct result of the improvement
shown in Fig.\ \ref{fig:AbsorptionSpectra} and previously
discussed. Alongside the quantitative data shown in
Fig.\ \ref{fig:AbsorptionSpectra}, this demonstrates that the crystal
spectral warping method, for a reasonable increase in computational
cost, can significantly improve on the accuracy of the hybrid dimer
calculations of absorption spectra and colour when compared to
experiment.

In Fig.\ \ref{fig:AbsorptionSpectraRangeSep}, we compare our crystal
spectral warping method to the results of treating a crystal supercell
of the YN polymorph fully with the hybrid functional PBE0, using
CP2K. Again, the same Gaussian broadening is applied as in
Fig.\ \ref{fig:SpecWarpAbsSpectra}. As previously mentioned in Section
\ref{subsec:CompDetails}, these hybrid crystal TDDFT calculations
were performed on a smaller supercell than the semi-local crystal
TDDFT calculations ($3\times2\times1$ versus $4\times3\times2$). This
should be borne in mind when comparing the results of the
calculations, but the results still give an estimate of the relative
accuracy of the crystal spectral warping method compared to full
hybrid calculations, and also demonstrate the limits of what accuracy
is achievable with limited computational time or hardware.  The PBE0
functional performs reasonably well, predicting a frequency $0.16$~eV
above the experimental peak, but the crystal spectral warping method
still slightly outperforms it. This is likely due to the larger
supercell size used in the semi-local ONETEP calculations, allowing
slightly longer-range effects to be included in the calculation. This,
coupled together with the significant reduction in cost, makes the
crystal spectral warping method superior in this case.  A further
comparison against calculations using a non-optimally tuned version of
the OT-SRSH functional\cite{manna_quantitative_2018} is presented in
Section S3 of the supporting information.

We also cross-check that the choice of code does not significantly
affect accuracy, as both CP2K and ONETEP are used in this work for
supercell calculations, by calculating the absorption spectrum of a
$3\times2\times1$ supercell with both codes, using PBE and placing no
localisation restriction on the excitations (results not shown). There
is a less than $1$\% difference in the frequency of the strongest
excitation, demonstrating that the choice of code does not affect the
result.

\section{Potential future extensions} \label{sec:Extensions}

The crystal spectral warping approach applied in this work is, at its
core, quite a general one, and there are a number of different avenues
through which it could be extended to include more complex
effects. Although we do not do include these effects in the present
work, in the following we discuss how the crystal spectral warping
method could be modified to include them.

One phenomenon would be the effect of vibrations and temperature on
the absorption spectrum. This is a topic that has been explored for
chromophores in solvent with
success\cite{zuehlsdorff_predicting_2017,zuehlsdorff_combining_2018}. To
include these effects, there are two main methods: the ensemble
approach, and the Franck-Condon approach, although these can be
employed simultaneously. The ensemble method typically includes the
motion of the nuclei at a classical level, by averaging the absorption
spectrum over a set of snapshots taken from a molecular dynamics
trajectory. In a solvated small-molecule system, on the order of
$100$-$1000$ snapshots are usually required for
convergence\cite{zuehlsdorff_predicting_2017,zuehlsdorff_combining_2018},
to ensure that configurational space is sufficiently sampled. The
number of snapshots required for a molecular crystal system such as
ROY would require similarly careful convergence. Assigning and keeping
track of symmetry-inequivalent dimers provides an additional
complication, but likely only a minor one.

Whilst the ensemble method includes low-frequency, classical,
vibrational contributions to the absorption spectrum, the
Franck-Condon approach aims to include quantum mechanical vibronic
effects, where electronic and vibrational degrees of freedom are
excited simultaneously\cite{zuehlsdorff_combining_2018}. This requires
the computation of overlap integrals between nuclear wavefunctions in
the ground and excited electronic states (under a set of simplifying
assumptions), and therefore an optimised initial geometry. Combining
the Franck-Condon and ensemble approaches na{\"i}vely, by computing
overlap integrals for each snapshot, is likely to be unfeasible
computationally, but work on solvated systems suggests that combining
the zero-temperature Franck-Condon result with the ensemble approach
gives good results\cite{zuehlsdorff_combining_2018}.

Another way to extend the crystal spectral warping method would be
through using methods other than semi-local/hybrid TDDFT for the two
levels of calculation. In particular, going beyond hybrid TDDFT could
allow phenomena such as excitons to be included; such methods could
include the Bethe-Salpeter
equation\cite{sagmeister_time-dependent_2009}, quantum Monte
Carlo\cite{feldt_excited-state_2020}, or even (multi-reference)
coupled cluster theory\cite{serrano-andres_calculation_2012}. However,
there are three considerations regarding which methods could be picked.

Firstly, it must be computationally feasible to use the two levels of
theory on the system sizes required of them -- it must be feasible to
apply the lower level of theory (semi-local TDDFT in this work) to
both small- and large-scale systems (in this work, dimers and
supercells respectively), and it must be feasible to apply the higher
level of theory (hybrid TDDFT in this work) to small-scale systems.

Secondly, the lower level of theory must be trusted to give a
qualitatively correct description of the system in both small- and
large-scale systems. If this is not true of the lower level of theory,
the crystal spectral warping method will not work, no matter which
theory is used for the higher-level calculations.

Thirdly, mapping excitations computed at the lower level of theory to
those computed at the higher level may become more complicated if the
two methods are less similar than in this work -- for example, if the
lower level of theory is density-based, but the higher level is
wavefunction-based. This is not a restriction, but does provide an
additional complication that would need to overcome.

\section{Conclusions} \label{sec:Conclusions}

In this work, we have presented the crystal spectral warping method,
based on applying a simple warping to large-scale crystalline
semi-local TDDFT calculations, to accurately predict the absorption
spectra of molecular crystals. This method is adapted from a method
previously used successfully for molecules in
solvent\cite{ge_accurate_2015,zuehlsdorff_predicting_2017}. We have
applied our crystal spectral warping method to three of the polymorphs
of the highly colour polymorphic compound ROY, and compared its
performance to a method based on hybrid TDDFT calculations on dimers,
and a theoretically more accurate but significantly more
computationally expensive full hybrid TDDFT method. The results of
these comparisons show that the crystal spectral warping method
improves on the hybrid dimer calculations, and also outperforms the
full hybrid method, at least for the PBE0 hybrid functional. Our
results also suggest that to calculate absorption spectra for
molecular crystals, different methods will be appropriate in different
scenarios, in order to balance accuracy and computational cost. If
high accuracy is required, but computational resources are limited,
the crystal spectral warping method presented here is likely to be the
best option. If only an approximate result is required, smaller hybrid
dimer TDDFT calculations may be more cost-effective.

This work presented here will allow significantly higher accuracy
calculations of absorption spectra of molecular crystals than was
possible previously, for a modest increase in computational expense,
making such calculations routine. The crystal spectral warping method
can also be easily implemented in a workflow, making it possible to
automate the prediction of absorption spectra. This should enable
high-throughput exploration of the optical absorption properties of
molecular crystals and their polymorphs, making the search for systems
with specific optical properties significantly more computationally
efficient.

\begin{suppinfo}

  Further details of pseudopotentials and basis sets used in this
  work; discussion of the effect of dispersion corrections on
  structure; comparison of results for the YN polymorph against
  non-optimally tuned range-separated hybrid functional results;
  outline of colour prediction theory used in this work; further
  details of the spectral warping parameters used in this work;
  discussion of the character of the excitations in dimer and
  supercell geometries; analysis of approximations made, including
  supercell size convergence, use of dimer as a basic unit, Gaussian
  basis set convergence, and use of the Tamm-Dancoff approximation.
    
\end{suppinfo}

\begin{acknowledgement}
  The authors acknowledge support from the UK Collaborative
  Computational Project for the Study of the Electronic Structure of
  Condensed Matter (CCP9) and the Engineering and Physical Sciences
  Research Council in the form of Software Infrastructure Grant
  EP/P02209X/1. The authors also acknowledge the support of the Thomas
  Young Centre for Theory and Simulation of Materials, through grant
  TYC-101. J.C.A.P.\@ acknowledges the support of St Edmund Hall,
  University of Oxford, through the Cooksey Early Career Teaching and
  Research Fellowship. Via our membership of the UK's HPC Materials
  Chemistry Consortium, which is funded by EPSRC (EP/L000202), and the
  UKCP consortium, which is also funded by EPSRC (EP/P022561/1), this
  work used the UK Materials and Molecular Modelling Hub (MMM Hub) for
  computational resources, which is partially funded by EPSRC
  (EP/P020194/1 and EP/T022213). The authors are also grateful to the
  Imperial College Research Computing Service for computational
  resources.\cite{imperial_hpc} The authors would like to acknowledge
  the use of the University of Oxford Advanced Research Computing
  (ARC) facility in carrying out this work.\cite{oxford_hpc} Finally,
  the authors would like to thank Dr.\@ Rui Guo for drawing our
  attention to this interesting system.
\end{acknowledgement}


\bibliography{ROYBib}

\end{document}


\title{Supporting Information for \\ Accurate and efficient computation of
  optical absorption spectra of molecular crystals: the case of the
  polymorphs of ROY}

\author{Joseph C.\ A.\ Prentice and Arash A.\ Mostofi}

\maketitle

\renewcommand\thefigure{S\arabic{figure}}
\renewcommand\thesection{S\arabic{section}}
\renewcommand{\bibnumfmt}[1]{S#1}
\renewcommand{\citenumfont}[1]{S#1}

\section{Pseudopotentials and NGWFs} \label{sec:PspotsAndNGWFs}
  
The pseudopotential strings defining the pseudopotentials used in
our CASTEP and ONETEP calculations were:
\begin{itemize}
\item H: \texttt{1|0.8|14|16|19|10N(qc=8)}
\item N: \texttt{1|1.1|23|26|31|20N:21L(qc=9)}
\item O: \texttt{1|1.2|23|26|31|20N:21L(qc=9)}
\item S: \texttt{1|1.8|6|7|8|30N:31L:32N}
\end{itemize}

ONETEP uses a basis set of localised non-orthogonal generalised
Wannier functions (NGWFs) to describe the electronic
structure. These functions are strictly localised, meaning they are
zero outside a given radius from the atom they are associated
with. This radius can be different for different atoms, and is a key
parameter that controls the quality of the basis set, and thus the
quality of the calculation. Different sets of NGWFs are used to
describe the valence and conduction states, as conduction states are
often more delocalised than valence states, and therefore require
larger radii. In the calculations presented here, the radii of the
valence NGWFs were chosen to be $8$~bohr for H and $9$~bohr for
other species. The conduction NGWF radii were chosen to be $10$~bohr
for H and $11$~bohr for other species.

\section{Inclusion of dispersion corrections} \label{sec:Dispersion}

\begin{table*}[t]
  \centering
  \begin{tabular}{ | c | c c c c |}
    \cline{2-5}
    \multicolumn{1}{ c }{} & \multicolumn{4}{ | c | }{$\theta$ ($^\circ$)} \\
    \hline
    Polymorph & Experiment\cite{yu_polymorphism_2010} & PBE & PBE+TS & PBE+TS (latt.\ params) \\
    \hline
    R & $34.5$ & $31.5$ & $30.6$ & $30.0$ \\
    ON & $44.7$ & $36.5$ & $41.4$ & $34.9$ \\
    YN & $58.8$ & $58.7$ & $62.7$ & $55.9$ \\
    \hline
  \end{tabular}
  \caption{Values of $\theta$ resulting from geometry optimisations
    with different levels of theory, as described in the main text. In
    the PBE and PBE+TS geometry optimisations, the lattice parameters
    are kept fixed at the value measured in
    experiment\cite{yu_polymorphism_2010}, with only atomic positions
    optimised. In the PBE+TS (latt.\ params) geometry optimisations,
    both the atomic positions and lattice parameters were optimised.}
  \label{tab:ThetaComparison}
\end{table*}

In this work, we use experimental structures obtained using X-ray
spectroscopy as a starting point. Such experimental measurements will
likely produce accurate values for large-scale structural parameters
of the system, such as the lattice parameters or the dihedral angle
$\theta$ defined in Fig.\ 1 in the main text; however, there will also
likely be significant errors in smaller-scale features, such as the
positions of light atoms (e.g. hydrogen). We would like to remove or
reduce these errors, whilst fixing the values of important large-scale
features at their experimental values. If the lattice parameters were
the only large-scale parameters we wanted to fix in this way, this
would be very simple -- we could simply perform a semi-local DFT
geometry optimisation, with the lattice parameters fixed at their
experimental values. Using the experimental lattice parameters also
allows us to implicitly include thermal expansion effects. However,
fixing $\theta$ is more difficult, as this constitutes a non-linear
constraint on the optimisation, which is very difficult to implement
practically. Instead, we aim to get the value of $\theta$ as close as
possible to experiment by comparing the result of optimising the
structures with and without the explicit inclusion of dispersion
interactions.

Including dispersion interactions, such as van der Waals forces, in
geometry optimisations of molecular crystals will most strongly affect
the lattice constants, as it affects how tightly molecules are bound
together. This will then have an indirect effect of the value of
$\theta$. By fixing the lattice constants at their experimental
values, as we do in this work, we implicitly include a large part of
this effect. A scheme for including dispersion interactions explicitly
will then provide small corrections, indirectly affecting the value of
$\theta$. We checked for such corrections by comparing the value of
$\theta$ (as discussed in Section 2 in the main text) for four
structures: the experimental structure, the PBE-optimised structure
with fixed lattice parameters, the PBE+TS-optimised structure with
fixed lattice parameters, and the PBE+TS-optimised structure with
optimised lattice parameters. TS here refers to the
Tkatchenko-Scheffler scheme for including dispersion interactions
semi-empirically\cite{tkatchenko_accurate_2009}. The results can be
found in Table \ref{tab:ThetaComparison}. We can see that, when the
lattice parameters are fixed at the experimental values, PBE alone
provides a better match to experiment than PBE+TS in two of the three
polymorphs -- R and YN. In the ON polymorph, PBE+TS gives a better
match to experiment than PBE alone. One potential explanation for this
result is the combination of the fact that explicitly including
dispersion interactions is expected to only provide small corrections
to the semi-local DFT result with experimental lattice parameters,
alongside the fact that the TS scheme is a simple semi-empirical
scheme, and thus will contain errors. If these errors are comparable
in size to these small corrections, it is quite possible that using
the TS scheme will actually give slightly worse agreement with
experiment than semi-local DFT alone.

In all three polymorphs, $\theta$ is poorly predicted when the lattice
parameters are optimised with PBE+TS too. Such an optimisation also
changes the lattice parameters significantly -- $a$ shrinks by
$12.1$\%, $2.7$\%, and $4.3$\% in YN, R, and ON respectively. This is
likely mostly due to the fact that the experimental lattice parameters
include the effect of thermal expansion, whereas the PBE+TS optimised
values are effectively at zero temperature. Taking the PBE+TS
optimised values for the lattice parameters as an approximation for
the true zero-temperature values, the change between these and the
experimental room-temperature values imply linear thermal expansion
coefficients of $4.59\times10^{-4}$, $0.92\times10^{-4}$, and
$1.50\times10^{-4}$~K$^{-1}$ for YN, R, and ON, respectively. These
values are comparable to those seen in other molecular
crystals\cite{ko_thermal_2018}, demonstrating that thermal expansion
likely accounts for the majority of the difference between the
experimental and PBE+TS lattice parameters.

The results of Table \ref{tab:ThetaComparison} demonstrate that the R
and YN polymorphs are best described using the PBE-optimised
structure, whilst the ON polymorph is best described using the
PBE+TS-optimised structure. They also show that ensuring the inclusion
of thermal effects on the structure via fixing the lattice parameters
at their experimental values is also important in obtaining an
accurate description.

\section{Comparison against (non-optimally tuned) OT-SRSH functional} \label{sec:OT-SRSH}

In addition to the comparison of the crystal spectral warping method
against supercell PBE0 TDDFT calculations for the YN polymorph in
Section 4 of the main text, we also compared our results against
supercell hybrid TDDFT calculations using a non-optimally tuned
version of the OT-SRSH functional defined in
Ref.\ \citenum{manna_quantitative_2018}. As our aim is to compute the
optical absorption spectrum both accurately and efficiently, we do not
specifically tune the parameters of the OT-SRSH functional to the
system in question (making the `OT' part of the name a slight misnomer
in this work). They were instead selected to be in the middle of the
range appropriate for similar molecular crystal systems:
$\alpha=0.2$\cite{manna_quantitative_2018}, $\beta=0.2444$, as
$\alpha+\beta=1/\epsilon_r$ and $n=\sqrt{\epsilon_r}\approx 1.5$ for
this kind of molecular crystal
\cite{jong_calculation_1991,jayatilaka_refractive_2009}, and
$\omega=0.25 \text{bohr}^{-1}$, as this lies in the middle of the
range found in other molecular
crystals\cite{manna_quantitative_2018}. It is worth noting that
$\alpha$ and $\beta$ here represent parameters of the OT-SRSH
functional,\cite{manna_quantitative_2018} and not the parameters of
the spectral warping transformation in Eq.\ (1) in
the main text.

These `OT'-SRSH calculations were performed in
CP2K\cite{kuhne_cp2k_2020} using the same supercell size and
parameters as the supercell PBE0 TDDFT calculations presented in
Section 4 in the main text. To implement the OT-SRSH
functional in CP2K, we included the following section in our input
file:
\texttt{ \\
  \&XC \\
    \&XC\_FUNCTIONAL NO\_SHORTCUT \\
      \&PBE \\
        PARAMETRIZATION ORIG  \# Use original PBE parametrization    \\ 
        SCALE\_X 0.5556 \# Always have at least (1-alpha-beta) fraction of PBE X \\
        SCALE\_C 1.0 \# Correlation is all PBE \\
      \&END PBE \\
      \&XWPBE        \# Short-range PBE functional \\
        SCALE\_X 0.2444 \# At short ranges, top up PBE X by beta \\
        SCALE\_X0 0.0 \\
        OMEGA 0.25  \# Should be the same as that for the HF potential \\
      \&END XWPBE \\
    \&END XC\_FUNCTIONAL  \\
    \&HF  \\
      \&SCREENING \\
        EPS\_SCHWARZ 1.0E-10 \\
        SCREEN\_ON\_INITIAL\_P .FALSE. \\
      \&END SCREENING  \\
      \&INTERACTION\_POTENTIAL \\
        CUTOFF\_RADIUS 5.8  \# Cutoff radius for truncated Coulomb interaction \\
        T\_C\_G\_DATA ./t\_c\_g.dat \\
        POTENTIAL\_TYPE MIX\_CL\_TRUNC \\
        SCALE\_COULOMB 0.2  \# Always have at least alpha fraction of HF X \\
        SCALE\_LONGRANGE 0.2444  \# At long ranges, top up HF X by beta  \\
        OMEGA 0.25   \# Should be the same as that for the XWPBE potential \\
      \&END INTERACTION\_POTENTIAL \\
    \&END HF \\
  \&END XC \\
}

\begin{figure}
  \centering
  \includegraphics[width=0.47\textwidth]{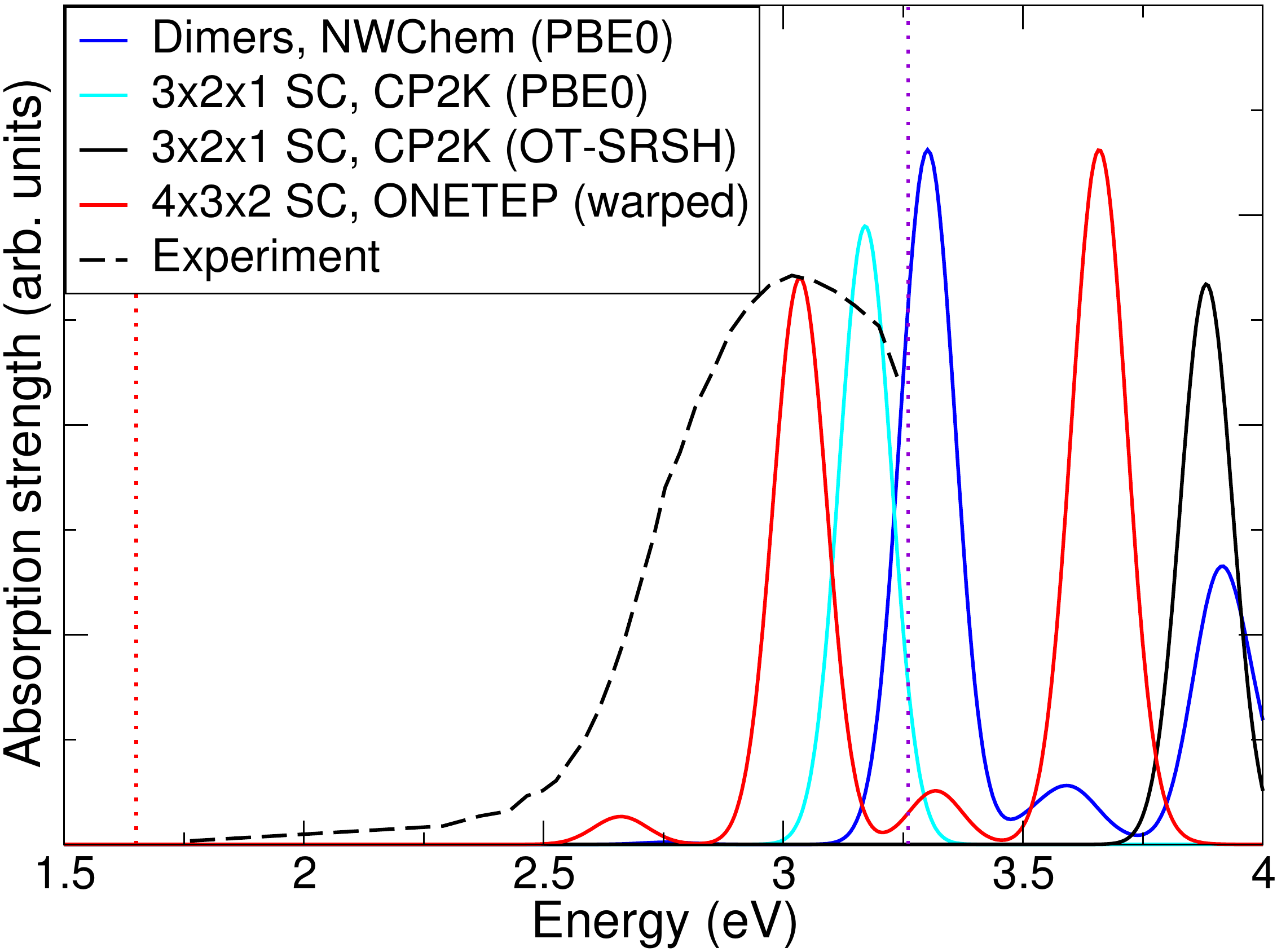}
  \caption{Absorption spectra for the YN polymorph calculated using
    several different methods. The data are the same as those
    presented in Fig.\ 7 in the main text, with the addition of the
    results of calculations performed with a non-optimally tuned
    version of the OT-SRSH functional. `\textit{$3\times2\times1$ SC,
      CP2K (PBE0)}' and `\textit{$3\times2\times1$ SC, CP2K
      (OT-SRSH)}' show the spectra calculated with CP2K for a
    $3\times2\times1$ supercell, using the relevant
    functionals. `\textit{$4\times3\times2$ SC, ONETEP (warped)}' and
    `\textit{Dimers, NWChem (PBE0)}' show the same data as
    `\textit{Semi-local crystal}' and `\textit{Hybrid dimer}' in
    Fig.\ 5c in the main text respectively, but have been renamed to
    emphasise the size of supercell and functional used. Experimental
    data is taken from Ref.\ \citenum{yu_color_2002}. The vertical
    dotted lines show the approximate edges of the visible spectrum.}
  \label{fig:AbsorptionSpectraRangeSepOTSRSH}
\end{figure}

In Fig.\ \ref{fig:AbsorptionSpectraRangeSepOTSRSH}, we additionally
compare the results already presented in Fig.\ 7 in the main text to
the results of treating a crystal supercell of the YN polymorph with
the version of the OT-SRSH functional described above. The results do
not compare particularly well against the crystal spectral warping
method or supercell PBE0 TDDFT results, predicting a frequency
$0.87$~eV above the experimental peak. This is likely because we have
not actually optimally tuned the parameters of the OT-SRSH functional,
but instead used values similar to those used in the literature. This
suggests that, although OT-SRSH functionals are capable of exceptional
accuracy, as in previous work\cite{manna_quantitative_2018}, they may
not be the best choice for general purpose calculations without first
undergoing the tuning process.

\section{Colour prediction theory} \label{sec:ColourTheory}

Given an optical absorption spectrum, such as that produced by a
TDDFT calculation, it is possible to calculate what colour a
material with this absorption spectrum would appear as to the human
eye. This can be done using tristimulus colorimetry
theory\cite{zuehlsdorff_predicting_2017,malcioglu_dielectric_2011,ge_accurate_2015}. We
first calculate the intensity of light transmitted through a sample
of the material as
\begin{equation}
  I(\lambda) = I_0(\lambda) e^{-\alpha(\lambda)t} ~,
\end{equation}
where $\alpha(\lambda)$ is the absorption spectrum from TDDFT as a
function of wavelength ($\lambda$), $t$ is the thickness of the
sample, and $I_0(\lambda)$ is the spectrum of the light source
illuminating the sample. The response of the cone cells in the human
retina to this spectrum must then be calculated. This is typically
done by using the tristimulus colour-matching functions
$x(\lambda)$, $y(\lambda)$, and $z(\lambda)$ to obtain the colour
indices $X$, $Y$, and $Z$:
\begin{equation}
  \begin{pmatrix} X \\ Y \\ Z \end{pmatrix} = N \int
  I(\lambda) \begin{pmatrix} x(\lambda) \\ y(\lambda)
    \\ z(\lambda) \end{pmatrix} \, d\lambda ~.
\end{equation}
The $XYZ$ colour indices can then be transformed to give $RGB$
indices, denoting the colour that would be observed.

In this work, we make use of the Colour Python
package\cite{colour_package} to do the bulk of this computation --
given the function $e^{-\alpha(\lambda)t}$ and the illuminant to use,
it can compute a colour swatch, like those seen in Fig.\ 6 in the main
text. We use the CIE standard D65
illuminant\cite{schanda_colorimetry_2007}, meant to approximate
daylight. For the generation of the function $e^{-\alpha(\lambda)t}$,
we need to determine an appropriate thickness $t$. For consistency, we
use the same thickness for the three calculated spectra for each
polymorph; we also use the same thickness for all three experimental
spectra. We take $t=\frac{A}{\bar{\alpha}_{\max}}$, where
$\bar{\alpha}_{\max}$ is the mean of the maximum values of
$\alpha(\lambda)$ of the three relevant spectra. This ensures that
$\alpha(\lambda)$ is normalised close to the range $[0,1]$, but
maintains the relative strength of the absorption peaks from different
spectra. $A$ is a constant, which is taken to be $10$ for the
calculated spectra and $40$ for the experimental spectra. In addition
to this, to make the width of our calculated spectra resemble
experimental spectra more closely, we apply a larger Gaussian
broadening than seen in the main text in Figs.\ 4, 5, and 7. In these
figures, a Gaussian broadening with standard deviation $0.0544$~eV was
applied, so that separated peaks can be easily discerned. For the
generation of the colour swatches, a broadening of $0.2177$~eV was
applied to the ON and YN spectra, and a broadening of $0.3265$~eV was
applied to the R spectra.

\section{Spectral warping parameters} \label{sec:SpectralWarpingParams}

Table \ref{tab:SpecWarpParams} shows the spectral warping parameters
$\beta$ and $s$, as defined by Equations (1) and (2) in the main text,
as derived from dimer calculations, for each of the
symmetry-inequivalent dimers in each of the polymorphs considered in
this work. As mentioned in the main text, $\alpha$ is taken to be $1$
throughout.

\begin{table*}[t]
  \centering
  \begin{tabular}{ | c | c c c c | c c c c | c c | }
    \hline
    Polymorph & \multicolumn{4}{ | c | }{R} & \multicolumn{4}{ | c | }{ON} & \multicolumn{2}{ | c | }{YN} \\
    \hline
    Dimer & 1 & 2 & 3 & 4 & \multicolumn{4}{ | c | }{1} & 1 & 2 \\
    Excitation & \multicolumn{4}{ | c | }{1} & 1 & 2 & 3 & 4 & \multicolumn{2}{ | c | }{1} \\
    \hline
    $\beta$ (meV) & $685$ & $611$ & $668$ & $649$ & $811$ & $904$ & $683$ & $698$ & $581$ & $607$ \\
    $s$ & $1.12$ & $1.73$ & $1.69$ & $1.37$ & $6.60$ & $12.04$ & $2.56$ & $0.97$ & $1.90$ & $2.12$ \\
    \hline
  \end{tabular}
  \caption{Values of the spectral warping parameters $\beta$ and $s$,
    as defined in Equations (1) and (2) in the main text, used in this
    work. `\textit{Dimer}' labels the various symmetry-inequivalent
    dimers within each polymorph -- spectral warping parameters are
    found for each dimer separately. `\textit{Excitation}' labels the
    (relevant) excitations seen in the absorption spectrum of each
    dimer -- each excitation is spectrally warped separately. Only the
    ON polymorph has more than one relevant excitation in the
    absorption spectrum, and the labels used for the excitations in
    this case match those used in Fig.\ 4 in the main text.}
  \label{tab:SpecWarpParams}
\end{table*}

\section{Character of the excitations} \label{sec:ExcitationCharacter}

\begin{figure*}[t]
  \centering
  \subcaptionbox{R Dimer}{
    \includegraphics[trim={8.5cm 2cm 8.5cm 2cm},clip,width=0.45\textwidth]{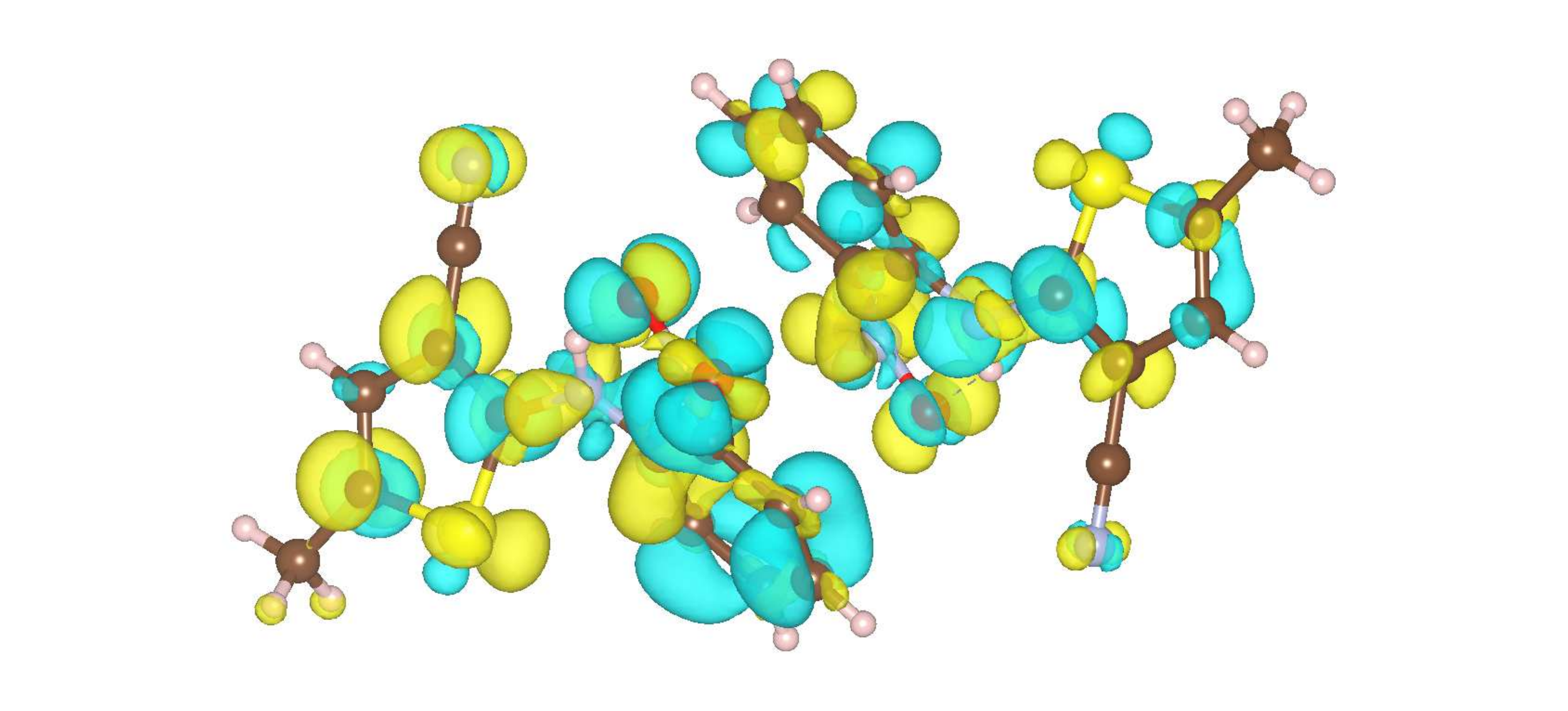}
  }
  ~
  \subcaptionbox{R $4\times3\times2$ supercell}{
    \includegraphics[trim={8.2cm 2cm 8.8cm 2cm},clip,width=0.45\textwidth]{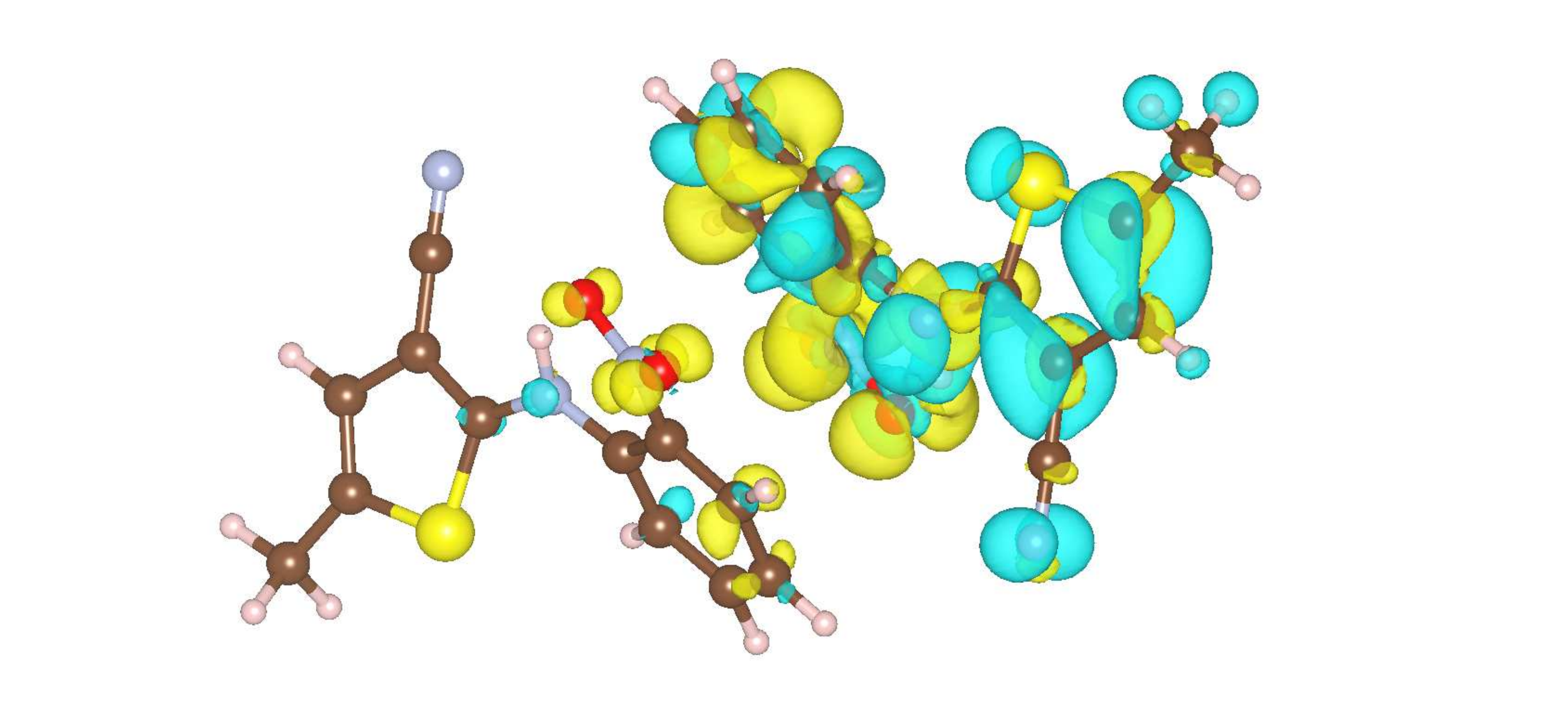}
  }
  
  \subcaptionbox{ON Dimer}{
    \includegraphics[trim={12cm 0cm 14cm 0cm},clip,width=0.45\textwidth]{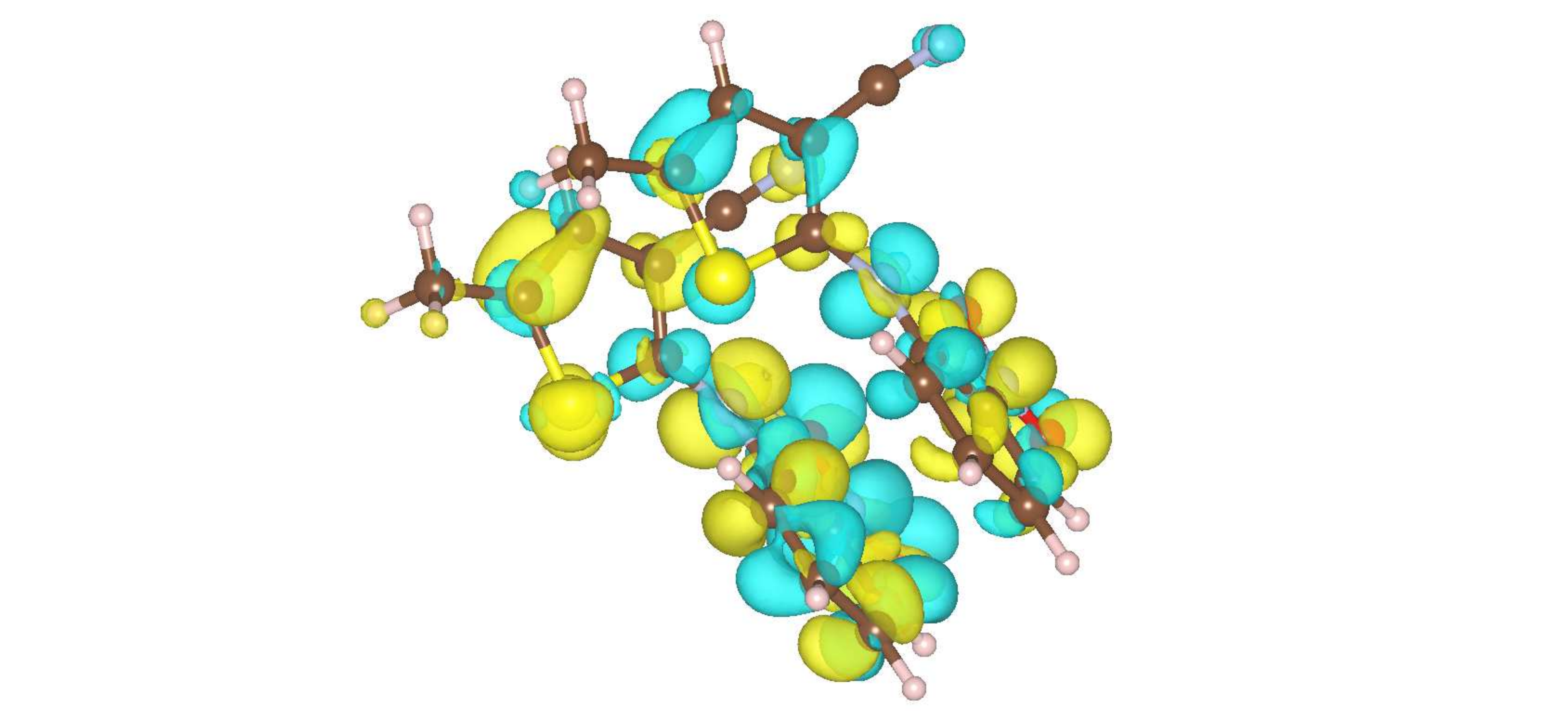}
  }
  ~
  \subcaptionbox{ON $4\times1\times2$ supercell}{
    \includegraphics[trim={11cm 1cm 16cm 1cm},clip,width=0.45\textwidth]{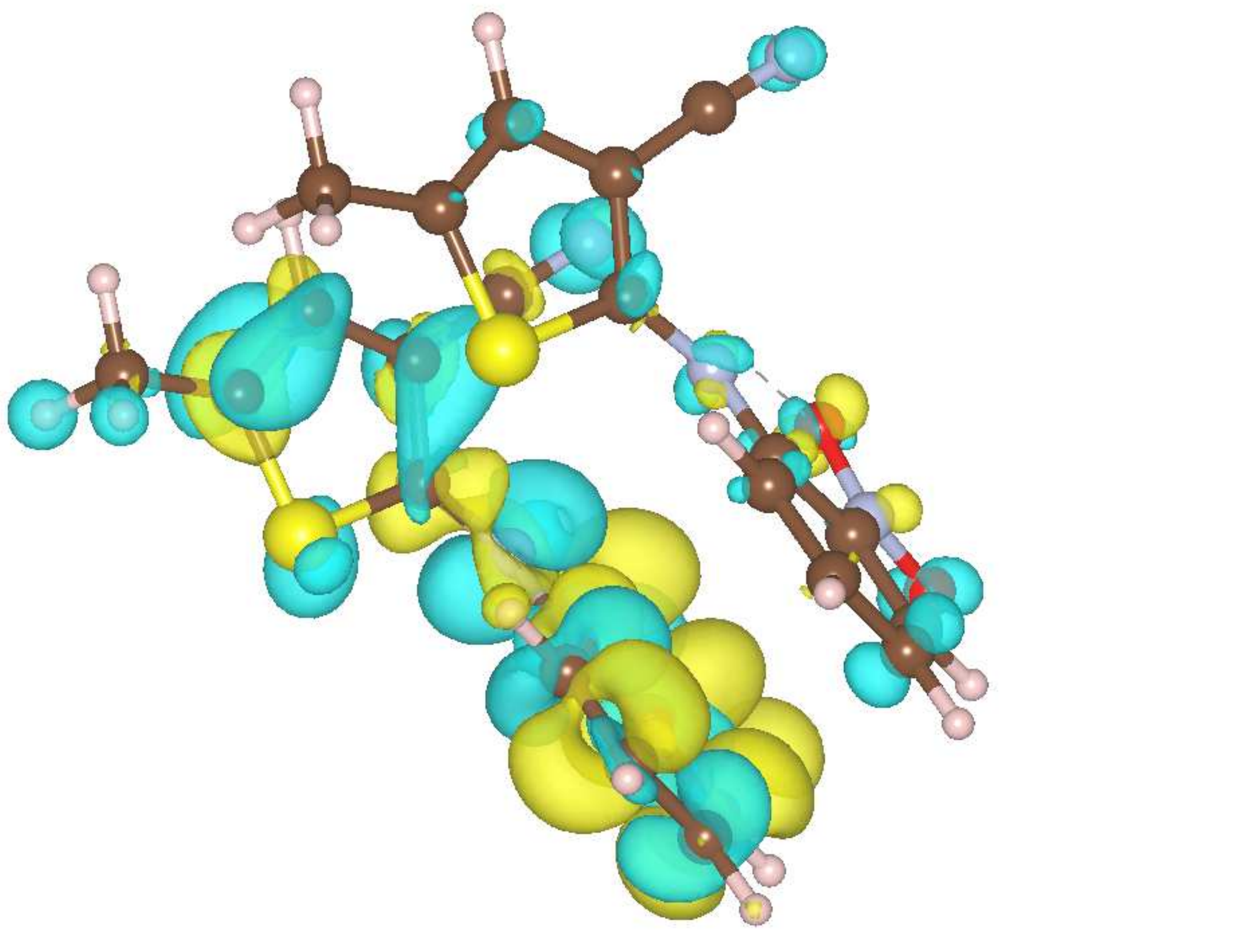}
  }
  ~
  \subcaptionbox{YN Dimer}{
    \includegraphics[trim={13cm 0cm 13cm 0cm},clip,width=0.45\textwidth]{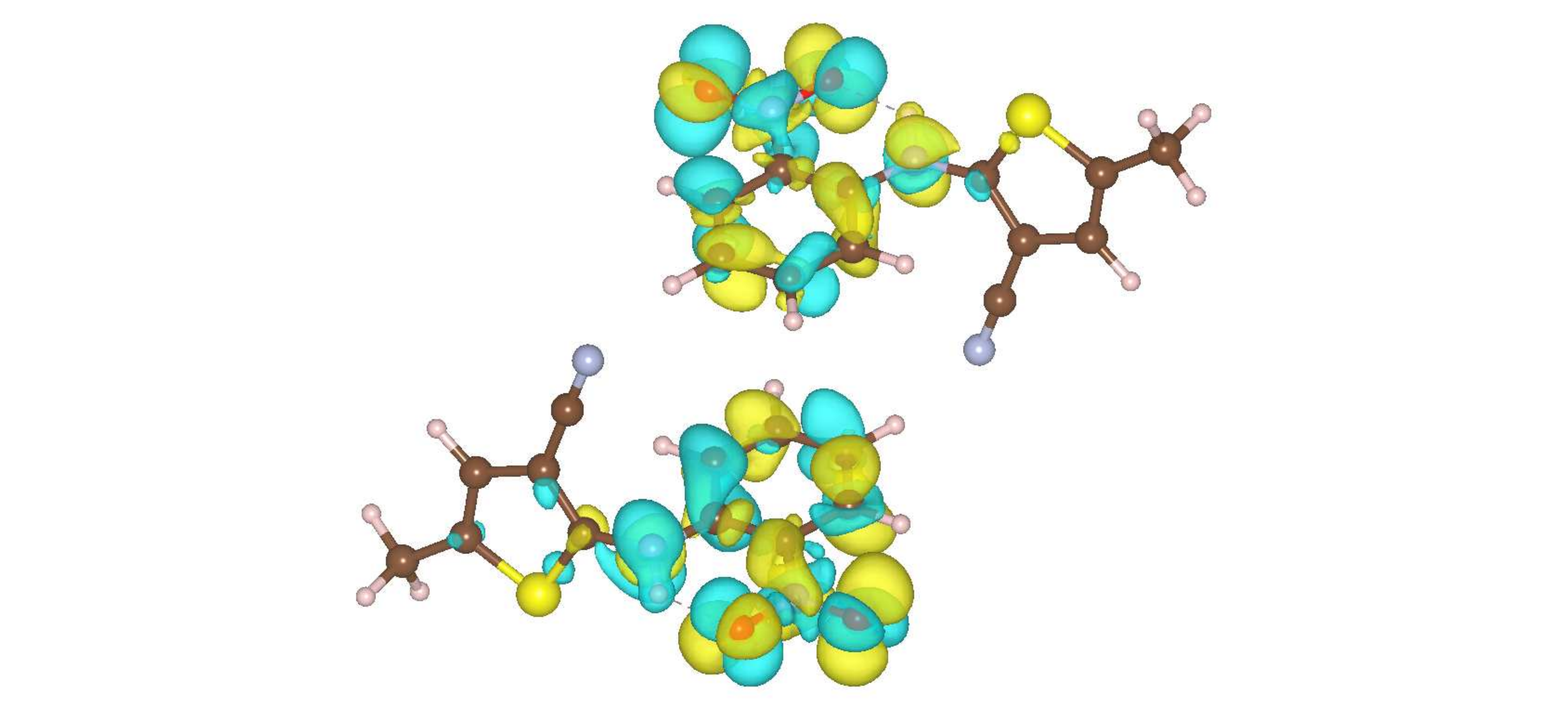}
  }
  ~
  \subcaptionbox{YN $4\times3\times2$ supercell}{
    \includegraphics[trim={13cm 0cm 13cm 0cm},clip,width=0.45\textwidth]{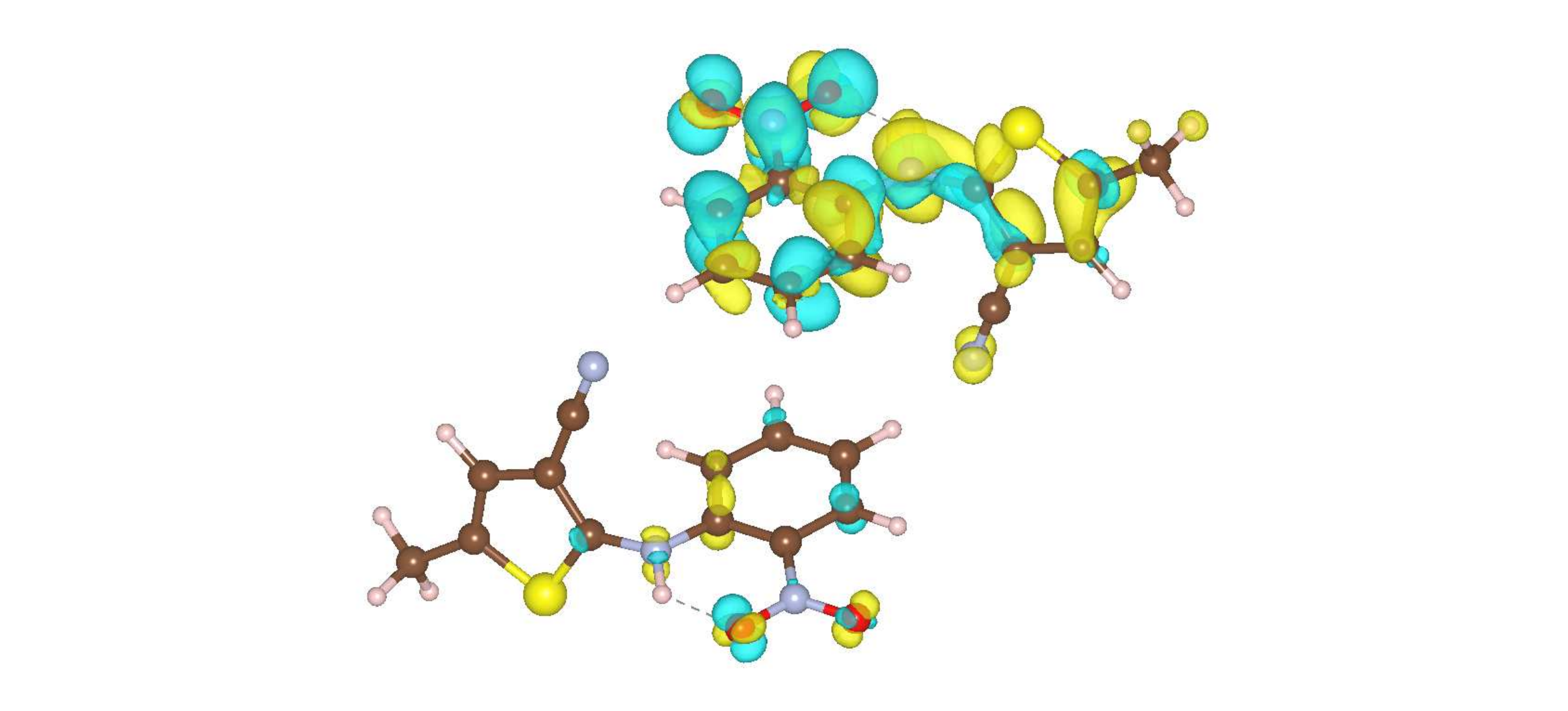}
  }
  \caption{Form of the strongest excitation in the R, ON and YN
    structures, calculated with ONETEP using PBE, as an isolated dimer
    or in a crystalline environment. The results for R for the dimer
    and crystalline environments are shown in (a) and (b)
    respectively, for ON in (c) and (d), and for YN in (e) and (f)
    respectively. The figures show an isosurface of the response
    density ($|n|=0.005$~e\,\r{A}$^{-3}$), with yellow and blue representing positive and
    negative response density values respectively. This isosurface is
    superimposed on the structure of the molecule. H, C, N, O, and S
    atoms are white, grey, blue, red, and yellow respectively. Figures
    produced using VESTA\cite{momma_vesta_2011}.}
  \label{fig:ExcitationCharacter}
\end{figure*}

The effect of the crystalline environment on the character of the
excitations can be investigated by considering the form of the
response density on a given dimer for the strongest excitation in
each polymorph, in both the isolated dimer and the crystalline
environment. To allow an exact comparison to be made, both the
crystalline and isolated dimer calculations were performed with
ONETEP, using the PBE functional. Open boundary conditions were used
for the dimer calculations. The results are shown in
Fig.\ \ref{fig:ExcitationCharacter}, which shows isosurfaces of the
response density for the strongest excitation in each structure
considered. In all three cases, the excitation changes its form
significantly under the influence of the crystalline
environment. The excitation becomes more localised onto one
molecule, instead of being spread over both, and the contribution of
the thiophenyl ring to the excitation increases (this is
particularly pronounced in the R and YN results). This increasing
localisation onto a single molecule may explain why single molecule
calculations are able to qualitatively predict the ordering of the
colours, as seen in Ref.\ \citenum{yu_color_2002}. 

\section{Analysis of approximations made}

\subsection{Supercell size convergence} \label{subsec:SupercellConvergence}

\begin{figure*}[t]
  \centering
  \includegraphics[width=0.45\textwidth]{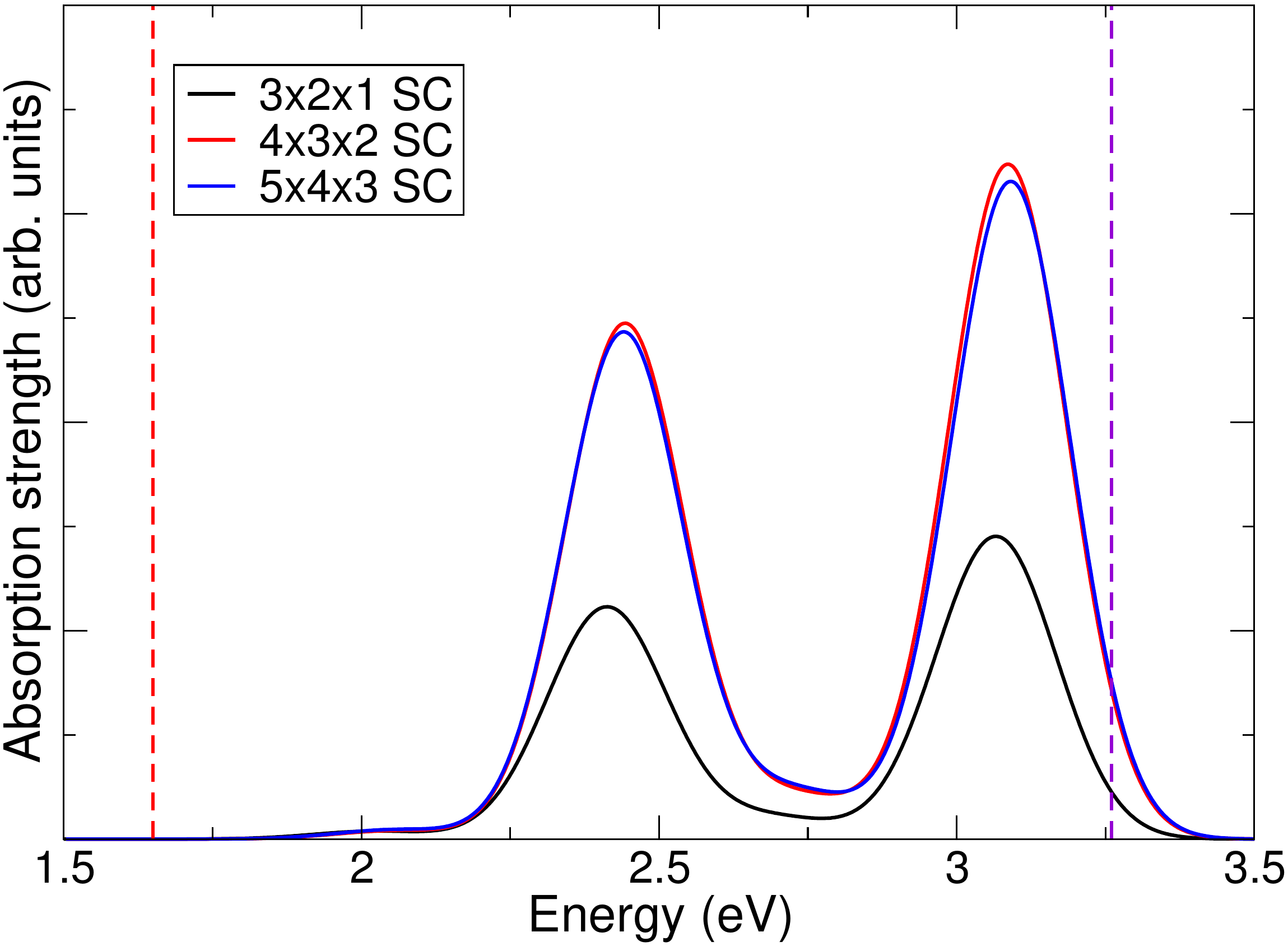}
  \caption{Comparison of the calculated absorption spectrum using
    supercells of increasing size ($3\times2\times1$,
    $4\times3\times2$, and $5\times4\times3$), for the YN
    polymorph. The calculations were performed using the PBE
    functional with ONETEP. The spectra shown are not a sum over
    symmetry-inequivalent dimers, as described in the main text, but
    are all calculated with the excitations restricted to be on the
    same dimer. The Tamm-Dancoff approximation was applied for all
    calculations. The vertical dotted lines show the approximate edges
    of the visible spectrum.}
  \label{fig:SupercellConvergence}
\end{figure*}

The crystal spectral warping method we apply in this work relies on
excitations being localised on a small number of molecules -- in the
system investigated here, a dimer. Without restricting the excitations
in this way, semi-local TDDFT would be unable to accurately describe
the system due to unphysical delocalisation. However, restricting the
excitations in a periodic TDDFT calculation also means that we need to
consider how to avoid spurious interactions between the excitations
and their periodic images, most likely through long-range
electrostatic dipole-dipole interactions. This issue only arises if
the excitation is restricted, and would be absent if the excitation is
allowed to delocalise over the whole system. To avoid these spurious
interactions, or at least make them small enough to neglect, it is
necessary to use a supercell, so that an excitation and its periodic
images are well separated. 

Fig.\ \ref{fig:SupercellConvergence} shows how the excitation spectrum
of the YN polymorph, with excitations restricted to a particular
dimer, changes as the size of the supercell increases from
$3\times2\times1$ to $4\times3\times2$ to $5\times4\times3$. Using the
$5\times4\times3$ supercell result as a reference, the excitation
energies for the lower energy peak in the $3\times2\times1$ and
$4\times3\times2$ calculations have an absolute error of $24$~meV and
$2$~meV respectively, whilst the corresponding oscillator strengths
have a relative error of $59$\% and $2$\% respectively. It is clear
that both quantities, particularly the oscillator strengths, exhibit
supercell size convergence. This shows, firstly, that the size of
supercell used in the calculations affects the results significantly,
and secondly, that the $4\times3\times2$ supercell provides an
excellent level of accuracy for the YN polymorph. Because of this, in
this work we use the $4\times3\times2$ supercell for the R and YN
polymorphs, and the similarly sized $4\times1\times2$ supercell for
the ON polymorph.

\subsection{Use of dimer as basic unit} \label{subsec:DimerConvergence}

\begin{figure*}[t]
  \centering
  \subcaptionbox{ON}{
    \includegraphics[width=0.45\textwidth]{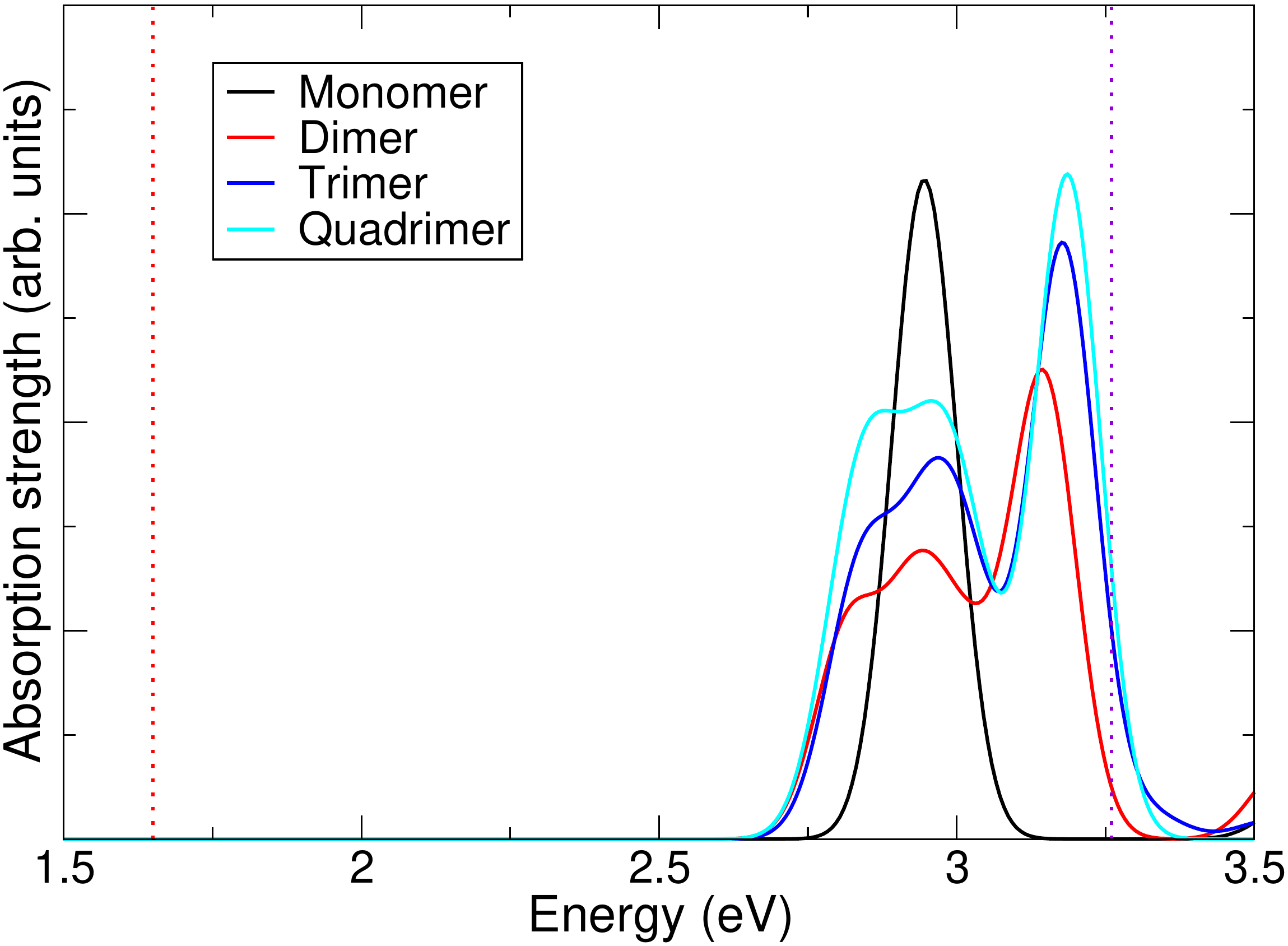}
  }
  ~
  \subcaptionbox{YN}{
    \includegraphics[width=0.45\textwidth]{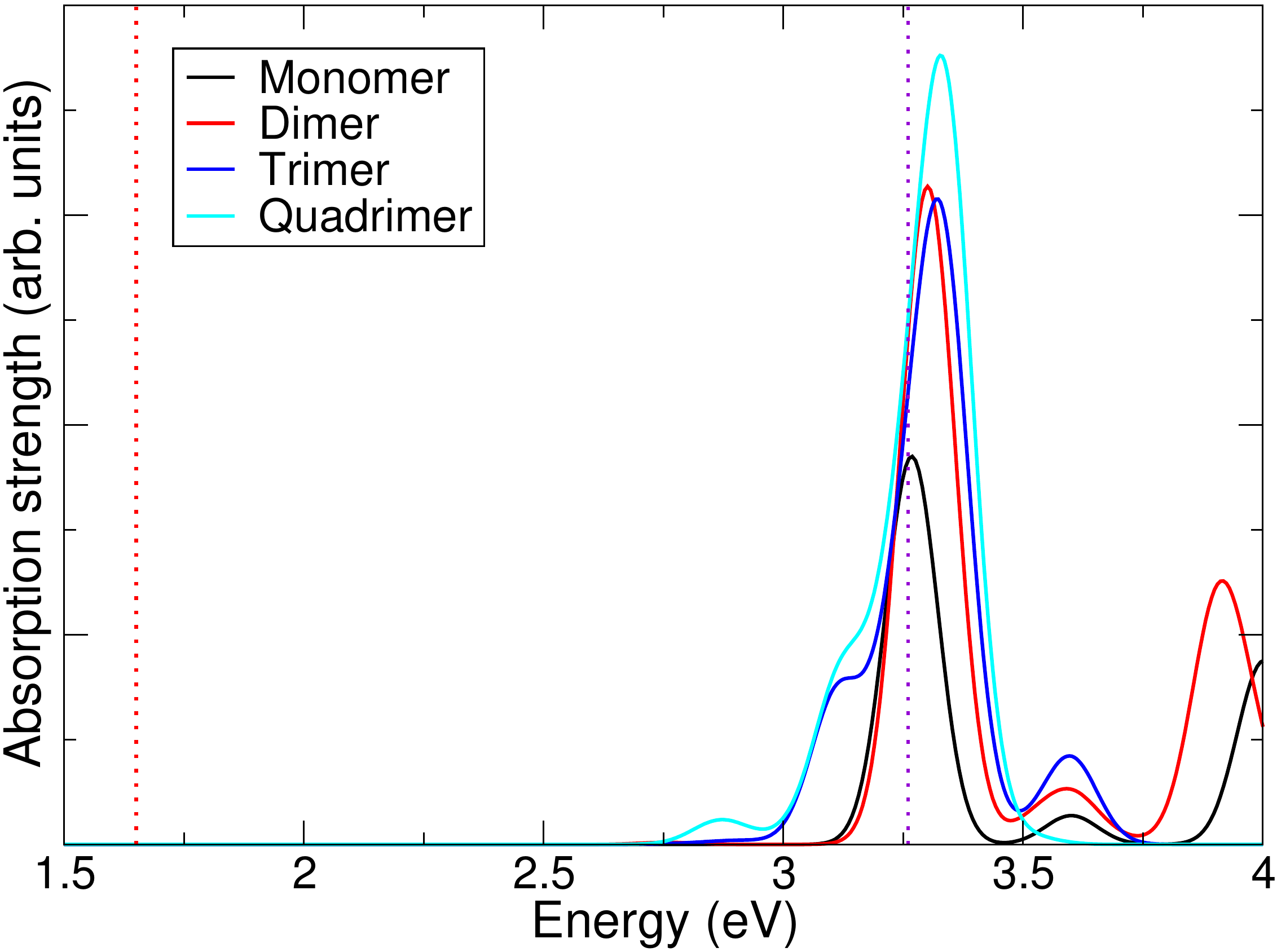}
  }
  \caption{Comparison of the calculated absorption spectrum for
    isolated molecular clusters of increasing size, for the ON and YN
    polymorphs. The clusters included are monomers, dimers, trimers,
    and quadrimers, containing 1, 2, 3, and 4 molecules
    respectively. The spectra shown are a sum over the spectra
    calculated for symmetry-inequivalent clusters, constructed by
    combining the symmetry-inequivalent dimers discussed in the main
    text. The results are calculated with
    NWChem\cite{apra_nwchem_2020} using the PBE0 functional and the
    cc-pVDZ basis set. The Tamm-Dancoff approximation was not
    applied. The vertical dotted lines show the approximate edges of
    the visible spectrum.}
  \label{fig:DimerConvergence}
\end{figure*}

One of the key assumptions of the (crystal) spectral warping method is
that we can qualitatively describe the relevant excitations of the
full (crystalline) system with a small isolated sub-system. This
sub-system should be small enough that we are able to perform both
semi-local and hybrid TDDFT on it, and thus calculate the spectral
warping parameters required to correct semi-local TDDFT calculations
performed on the full system. This is the process described in Section
3.2 of the main text. What this sub-system should include is,
therefore, an important question.

If our excitations are largely molecular in nature (which they are in
this work), the simplest sub-system to use would simply be a monomer
-- an isolated molecule. We can then go further, and consider the
effect of molecular interactions on electronic excitations. To a first
approximation, we would expect that interactions between nearest
neighbour molecules would have the strongest effect on the
excitations, suggesting using molecular dimers as the sub-system
should improve the accuracy significantly. More distant molecules will
likely have a smaller effect, suggesting that using molecular trimers
and quadrimers will provide smaller corrections to the calculated
electronic excitations. 

To save computational expense, we would like to use the smallest
sub-system that still provides an accurate qualitative description of
the absorption spectrum. We can test this by performing calculations
on several sub-system sizes, ranging from monomer to quadrimer, and
analysing the results. Figs.\ \ref{fig:DimerConvergence}a and b show
such results for the ON and YN polymorphs respectively, calculated
using the PBE0 functional. As in Figs.\ 4 and 5 in the main text, for
the YN polymorph the dimer, trimer, and quadrimer spectra shown are
actually a sum over symmetry-inequivalent dimers/trimers/quadrimers,
normalised by the number of dimers/trimers/quadrimers included in the
sum. The trimers and quadrimers considered were constructed by
combining the symmetry-inequivalent dimers. The results clearly show
that the dimer is the smallest sub-system that provides a
qualitatively accurate description of the absorption spectrum for both
polymorphs. In the ON polymorph, the dimer calculation correctly
describes the multiple-peaked structure of the spectrum, with the
energies of the peaks very close to the values in the larger quadrimer
calculation. In the YN polymorph, although the dimer calculation is
missing one or two small subsidiary peaks compared to the quadrimer
calculation, it accurately describes the energy of the main absorption
peak, which dominates the spectrum. In both polymorphs, using trimers
or quadrimers only provides small corrections to the dimer
results. This is consistent with these corrections being largely
environmental screening-like in nature. Since such effects are
included in the crystal spectral warping method via the semi-local
crystal TDDFT calculations, it is unnecessary to include them within
the unit that excitations are localised on.

These results therefore suggest that dimers are an appropriate choice
for the basic unit of our calculations, as they are the smallest
sub-system that is able to provide a qualitatively correct result
compared to larger clusters, whilst not including effects that will be
included at another point in our calculations.

\subsection{Use of cc-pVDZ basis set} \label{subsec:AugBasisComparison}

For all (TD)DFT calculations, it is important to ensure that the basis
set used is able to accurately describe the Kohn-Sham wavefunctions of
the system, and thus the density. In plane wave codes, such as CASTEP
(and ONETEP indirectly), this can be done in a systematic way, by
increasing the plane wave cut-off energy until the basis set is of
sufficient size. For codes that use Gaussian basis sets, such as
NWChem, this is more difficult. One way that Gaussian basis sets can
be `improved' in some cases is by the addition of diffuse basis
functions, that extend further from the atoms than other basis
functions. These can be useful for describing intermolecular
interactions\cite{papajak_perspectives_2011}. However, including
diffuse basis functions is often significantly more computationally
expensive, not only because they directly increase the complexity of
the calculation, but because they often cause problems with numerical
stability. For this reason, it is better to avoid the use of diffuse
basis functions unless required\cite{papajak_perspectives_2011}.

In this work, we use the cc-pVDZ basis set for all NWChem
calculations, which does not include diffuse basis functions. There is
also a version of this basis set with added diffuse basis functions,
aug-cc-pVDZ. Investigating the effect of including these diffuse basis
functions is important, as this could potentially affect the spectral
warping parameters, and thus the crystal spectral warping result. It
could also affect the hybrid dimer results that we compare our crystal
spectral warping results against.

\begin{figure*}[t]
  \centering
  \subcaptionbox{ON}{
    \includegraphics[width=0.45\textwidth]{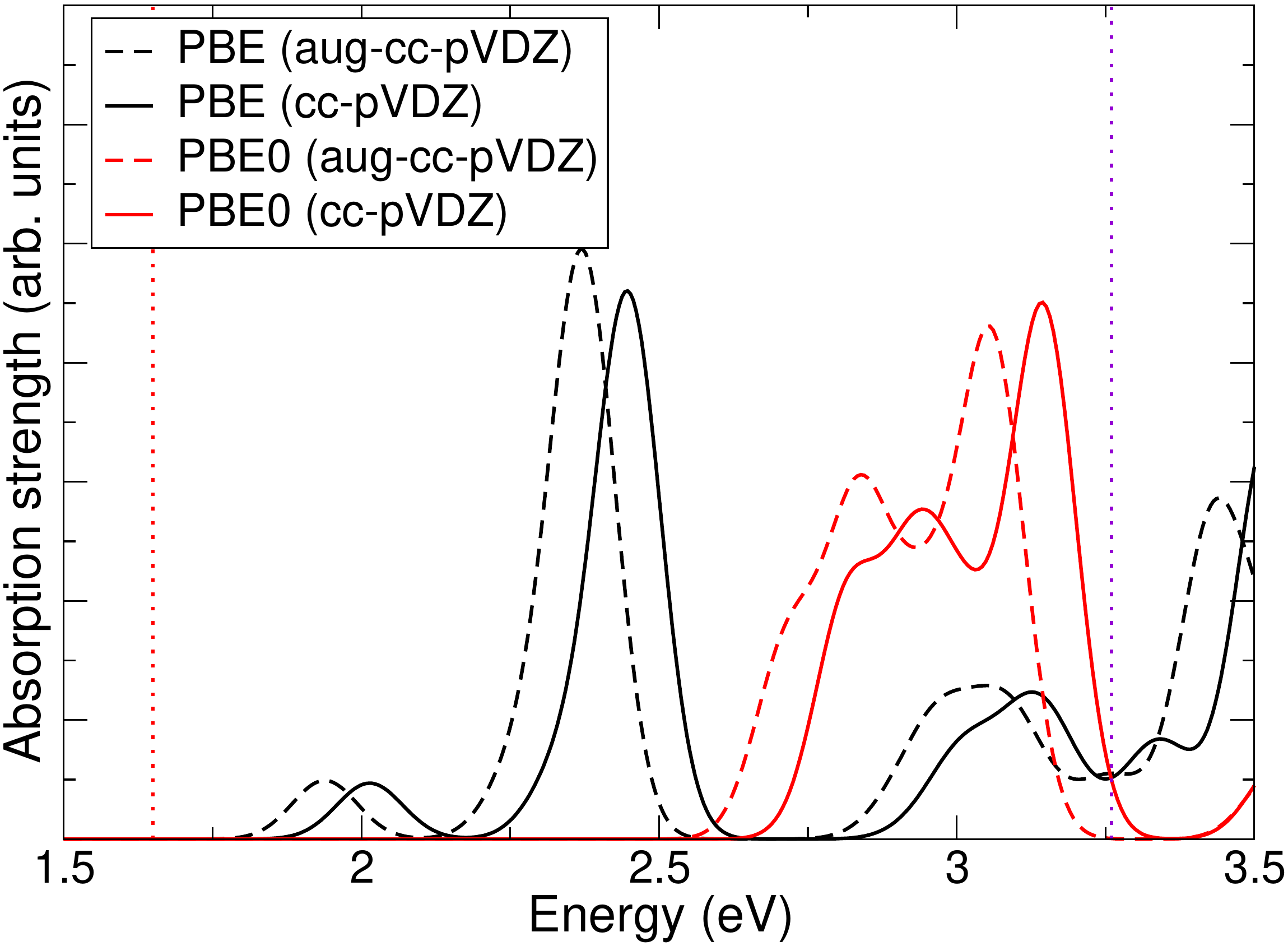}
  }
  ~
  \subcaptionbox{YN}{
    \includegraphics[width=0.45\textwidth]{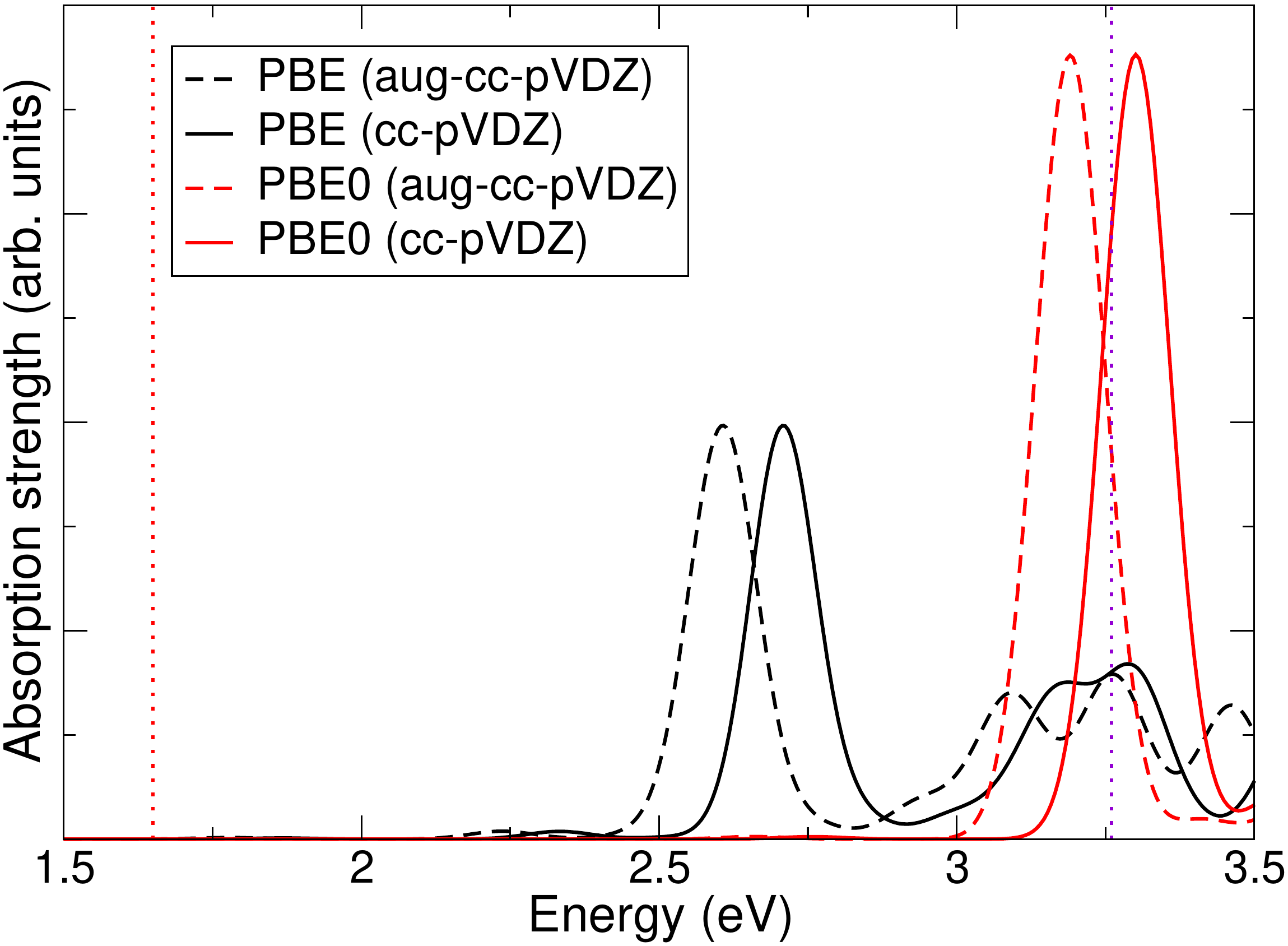}
  }
  \caption{Comparison of the dimer absorption spectrum, calculated
    using the cc-pVDZ basis set used in the rest of this work, and the
    aug-cc-pVDZ basis set, which includes diffuse functions, for the
    ON and YN polymorph. The spectra shown are a sum over
    symmetry-inequivalent dimers, as described in the main text. The
    results are calculated with NWChem\cite{apra_nwchem_2020} using
    the PBE and PBE0 functionals. The Tamm-Dancoff approximation was
    not applied. The vertical dotted lines show the approximate edges
    of the visible spectrum.}
  \label{fig:AugBasisComparison}
\end{figure*}

In order to compare the effect of including diffuse basis functions,
Fig.\ \ref{fig:AugBasisComparison} shows absorption spectra calculated
using both the cc-pVDZ and aug-cc-pVDZ basis sets for the ON, and YN
polymorphs, using the PBE and PBE0 functionals. The spectra are
superpositions of the calculated spectra for symmetry-inequivalent
dimers, as in Figs.\ 4 and 5 in the main text. We only present results
for two of the polymorphs because, as noted above, the aug-cc-pVDZ
calculations are significantly more computationally expensive than the
cc-pVDZ calculations. It can be seen that including diffuse basis
functions lowers excitation energies by roughly $0.1$~eV in all cases,
but that the form of the spectrum remains very similar. The rigidity
of this shift means that the spectral warping shifts $\beta$ change by
a mean of $13$~meV over both polymorphs, and a maximum of
$33$~meV. This shows that the spectral warping method is largely
unaffected by the inclusion of diffuse basis functions.

The key point to note here is that the spectral warping parameters by
looking at the difference between two TDDFT spectra, calculated at
different levels of theory. Because of this, there is some degree of
error cancellation, as errors that are present in both calculations
(e.g. due to the neglect of diffuse basis functions) will cancel
out. This allows accurate values for the spectral warping parameters
to be calculated when the individual spectra may contain systematic
errors. The results shown in Fig.\ \ref{fig:AugBasisComparison} show
that this is in fact the case here -- although we have a small error
in the cc-pVDZ spectra relative to the aug-cc-pVDZ spectra, this error
almost entirely cancels out when calculating the spectral warping
parameters. This demonstrates that using a basis set that doesn't
include diffuse basis functions, such as cc-pVDZ, is sufficiently
accurate for the purposes of this work. As this also significantly
saves on computational expense, we choose to neglect diffuse basis
functions throughout this work.

The effect of diffuse basis functions on the hybrid dimer results that
the crystal spectral warping results are compared to is discussed in
conjunction with the effect of the TDA in Section
\ref{subsec:TDAComparison}.

\subsection{Use of Tamm-Dancoff approximation} \label{subsec:TDAComparison}

\begin{figure*}[t]
  \centering
  \subcaptionbox{R}{
    \includegraphics[width=0.45\textwidth]{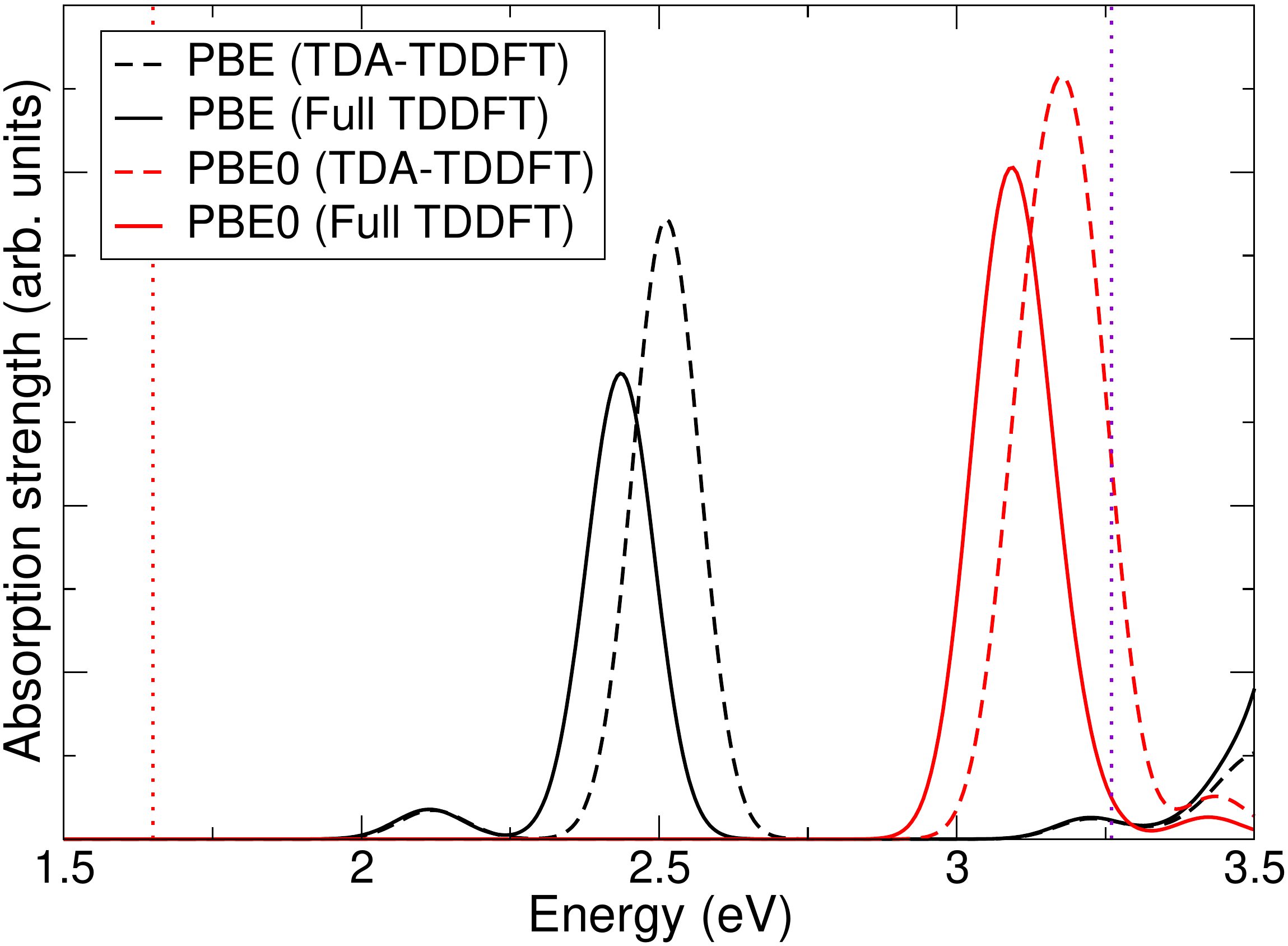}
  }
  ~  
  \subcaptionbox{ON}{
    \includegraphics[width=0.45\textwidth]{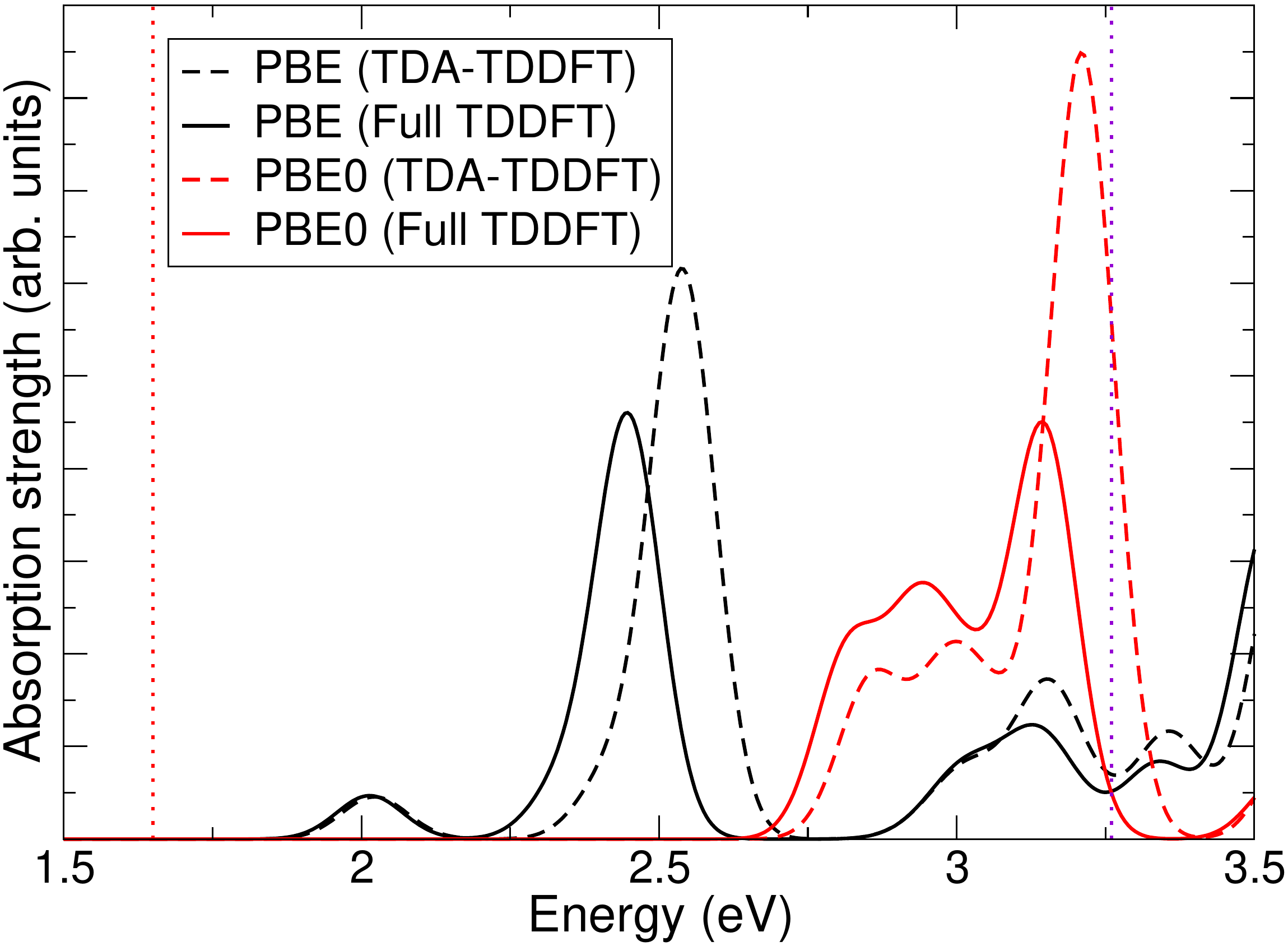}
  }
  ~
  \subcaptionbox{YN}{
    \includegraphics[width=0.45\textwidth]{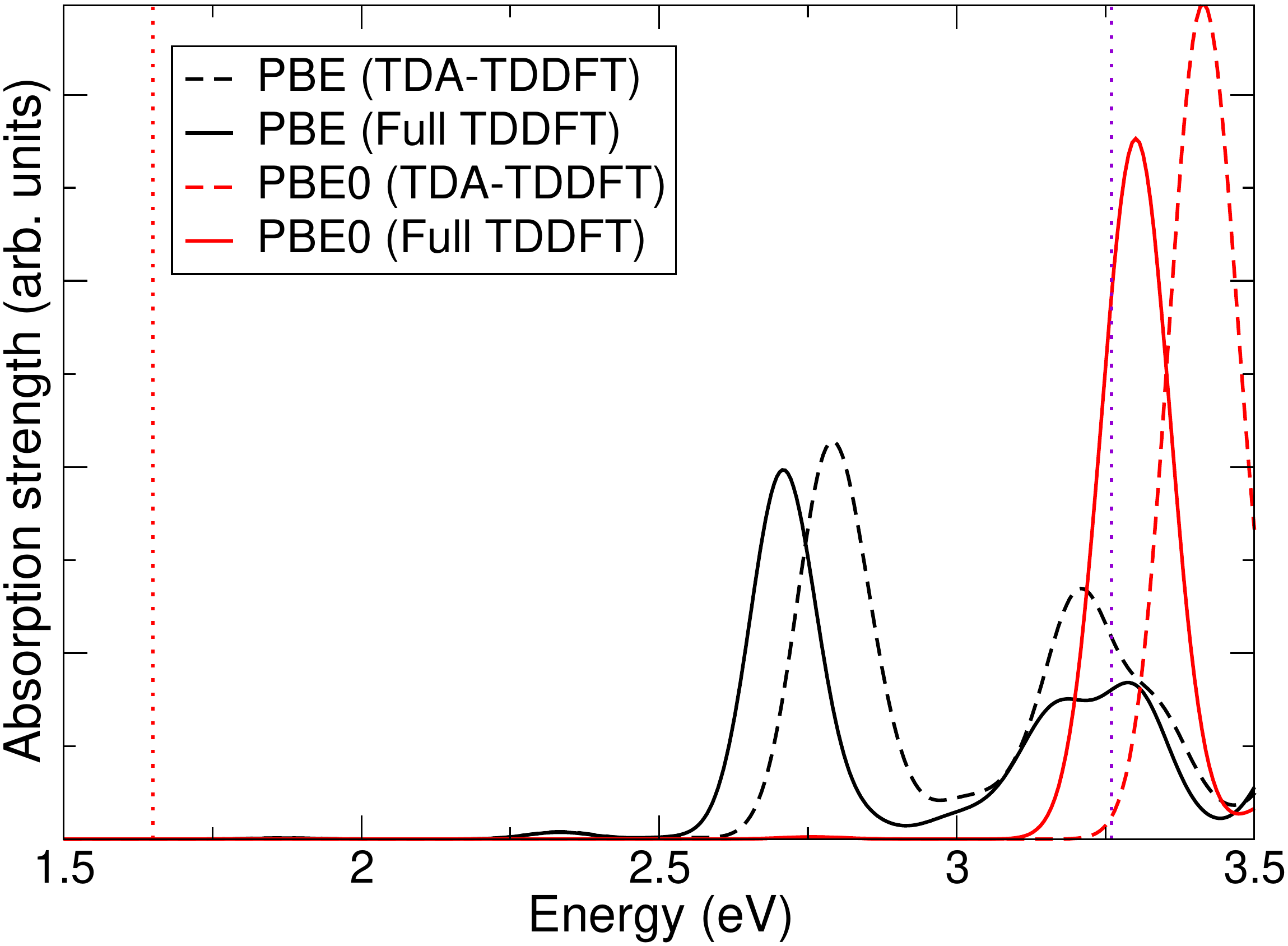}
  }
  \caption{Comparison of the dimer absorption spectrum, calculated
    with and without the Tamm-Dancoff approximation, for the R, ON,
    and YN polymorphs. The spectra shown are a sum over
    symmetry-inequivalent dimers, as described in the main text. The
    results are calculated with NWChem\cite{apra_nwchem_2020} using
    the PBE and PBE0 functionals and the cc-pVDZ basis set. The
    vertical dotted lines show the approximate edges of the visible
    spectrum.}
  \label{fig:TDAComparison}
\end{figure*}

As discussed in the main text, the Tamm-Dancoff approximation (TDA)
reduces the computational expense of TDDFT calculations, whilst
typically maintaining accurate values for the excitation energies,
although oscillator strengths can be less accurate. In this work, we
use the TDA for all calculations performed with ONETEP and CP2K,
whilst we use full TDDFT for all NWChem calculations. 

In order to compare the relative accuracy of the TDA and full TDDFT
methods, Fig.\ \ref{fig:TDAComparison} shows absorption spectra
calculated using both the TDA and full TDDFT for the R, ON, and YN
polymorphs, using the PBE and PBE0 functionals. The spectra are
superpositions of the calculated spectra for symmetry-inequivalent
dimers, as in Figs.\ 4 and 5 in the main text. It can be seen that
using the TDA raises excitation energies by roughly $0.1$~eV compared
to full TDDFT in all cases, but that the form of the spectrum remains
very similar. The rigidity of this shift means that the spectral
warping shift changes little -- the average change is $25$~meV across
all polymorphs, and the maximum change is $45$~meV. This suggests that
the spectral warping method is largely unaffected by whether TDA or
full TDDFT is used. There is a cancellation of errors, similar to that
described in Section \ref{subsec:AugBasisComparison} for diffuse basis
functions, that means that both the TDA and full TDDFT give very
similar values for the spectral warping parameters.

\bibliographystyle{unsrt}
\bibliography{ROYBib}